\documentclass[prb,aps,amssymb,bbm,twocolumn,showpacs,amsmath]{revtex4}

\usepackage{graphicx,bbm}

\begin{document}

\title{Dynamics of Coupled Qubits Interacting with an Off-Resonant Cavity}
\author{Oliver Gywat$^{1,2}$}
\author{Florian Meier$^{2}$}
\author{Daniel Loss$^1$}
\author{David~D. Awschalom$^{2}$}
\affiliation{$^1$Department of Physics and Astronomy, University
of Basel, Klingelbergstrasse 82, 4056 Basel, Switzerland \\
$^2$Center for Spintronics and Quantum Computation, University of
California, Santa Barbara, California 93106, USA}
\date{\today}

\begin{abstract}
We study a model for a pair of qubits which interact with a single off-resonant cavity mode and, 
in addition, exhibit a direct inter-qubit coupling. 
Possible realizations for such a system include coupled superconducting
qubits in a line resonator as well as exciton states or electron spin states of quantum dots in a  cavity. 
The emergent dynamical phenomena are strongly dependent on the relative energy scales of the inter-qubit coupling strength, the coupling strength between qubits and cavity mode, and the 
cavity mode detuning. 
We show that the cavity mode dispersion enables a measurement of the state of the coupled-qubit 
system in the perturbative regime. 
We discuss the effect of the direct inter-qubit interaction on a cavity-mediated two-qubit gate.
Further, we show that for asymmetric coupling of the two qubits to the cavity, the direct 
inter-qubit coupling can be controlled optically via the ac Stark effect.
\end{abstract}


\maketitle

\section{Introduction}
\label{sec:intro}

The quantum-mechanical coupling of individual atoms and photons
has been demonstrated in a series of cavity-QED experiments during
the past decade.~\cite{raimond:01,turchette:95,varcoe:00} In
addition to the fundamental interest in states formed by
superpositions of matter and photon degrees of freedom, the
development of cavity-QED has been triggered by possible
applications in quantum information schemes. Strong coupling of a
qubit and a cavity mode allows one to convert localized qubits into
flying qubits suitable for transmission of quantum
information.~\cite{enk:97} 
Transmission through a detuned cavity
mode can be utilized for qubit
readout.~\cite{blais:04,wallraff:04} For several qubits
interacting with a single cavity mode, the cavity can act as bus for
quantum information that couples qubits with large spatial
separation.~\cite{imamoglu:00,blais:04}

In this manuscript, we consider a pair of qubits both of which are coupled to one
cavity mode and, in turn, both qubits are coupled directly to one another by an
exchange-type interaction.
The Hamiltonian of the system under study is given by
\begin{eqnarray}
\hat{H} &=& \hbar  \sum_{i=1,2} \left[ \Omega_i \hat{s}_{i,z} +
g_i \left(\hat{a} \hat{s}_i^+ + \hat{a}^\dagger \hat{s}_i^- \right)\right] \label{eq:gen-ham} \\
&& + \hbar \omega \hat{a}^\dagger \hat{a} + \frac{\hbar
J_\perp}{2} \left( \hat{s}_1^+  \hat{s}_2^- + \hat{s}_1^-
\hat{s}_2^+ \right) + \hbar J_z \hat{s}_{1,z} \hat{s}_{2,z}.
\nonumber
\end{eqnarray}
Here, $\hbar \Omega_i$ is the energy splitting of qubit $i=1,2$,
$\hbar g_i$ is the coupling strength between qubit $i$ and the cavity mode, $\hbar
\omega$ is the energy of a cavity photon, $\hat{s}_i^\pm$ are raising and
lowering operators for qubit $i$, and $\hat{s}_{i,z}=(1/2)\sigma_{i,z}$ are 
pseudo-spin operators with the
Pauli matrix $\sigma_{i,z}$.  Extending
previous work~\cite{imamoglu:00,blais:04} on cavity-QED with solid-state
systems, we take into account an additional direct qubit coupling with
amplitudes $\hbar J_{\perp}$, $\hbar J_{z}$. The interplay of direct and cavity-mediated
coupling leads to intricate phenomena. For $J_z = J_\perp = 0$,
the state of both qubits can be read out simultaneously from the
cavity dispersion if $\Omega_1 \neq \Omega_2$ and the cavity
$Q$-factor is sufficiently large.~\cite{blais:04,wallraff:04} We
show that a similar readout scheme remains effective even for the
case of coupled qubits. For identical qubits, $\Omega_1 =
\Omega_2$ and $g_1=g_2$, the measurement via the cavity mode
projects the coupled state onto the singlet-triplet basis and thus
would allow one to detect entangled two-qubit states. We show that
the direct exchange coupling modifies the cavity-mediated
interactions and decreases the
fidelity of cavity-mediated two-qubit
gates.~\cite{imamoglu:99,blais:04} On a more fundamental level,
our model allows us to investigate the transition from
cavity-mediated inter-qubit interactions at small $J_{\perp}$ to a system of two directly
coupled qubits interacting with a cavity.

An implementation of the Hamiltonian Eq.~(\ref{eq:gen-ham}) is
feasible with various physical systems.
For
$J_{\perp}=J_z=0$, Eq.~(\ref{eq:gen-ham}) has been derived for
superconducting qubits at the charge degeneracy point interacting
with a microwave resonator.~\cite{blais:04} In the present work, this scheme is extended to capacitively coupled superconducting qubits.
Other implementations include exciton states in coupled quantum dots or coupled quantum shells 
in optical microcavities.~\cite{kiraz:03,vuckovic:03} We show that this model can readily be extended to electron spin states in coupled quantum dots,\cite{loss:98,burkard:99,burkard:00} interacting with a microwave cavity or an optical cavity.

This work is organized as follows.
In Sec.~\ref{sec:dispersive}, we calculate the optical response of the
coupled qubits in the dispersive regime, where the cavity mode is
far off resonance with all transitions taking place in the coupled-qubit
system. We focus on the case of identical and of non-identical qubits with finite direct coupling.
For identical qubits we derive the fidelity for a cavity-mediated two-qubit gate in the presence of the direct qubit interaction.
In Sec.~\ref{sec:disp-control}, we discuss the control of the direct qubit interaction via the ac Stark shift for qubits with $g_1 \neq g_2$.
In Sec.~\ref{sec:microscopic}, we
discuss possible experimental implementations of
Eq.~(\ref{eq:gen-ham}) in physical systems and derive explicit
expressions for the energy scales in Eq.~(\ref{eq:gen-ham}) for
quantum dots, quantum shells, and superconducting qubits in a cavity.
We conclude in Sec.~\ref{sec:conclusion}.

\section{Schrieffer-Wolff transformation of the Hamiltonian}
\label{sec:dispersive}

We consider the dispersive regime, where the cavity mode is non-resonant with all transitions of the qubit
system and the coupling of cavity and qubits in Eq.~(\ref{eq:gen-ham}) can be treated perturbatively. For a
single qubit, the cavity resonance was shown~\cite{blais:04} to shift by $\pm g^2/\Delta$, where $\Delta$ is the photon
detuning, depending on the state of the qubit. Measurement of the cavity resonance hence provides a readout
mechanism for the qubit which has been demonstrated for Cooper pair boxes at the charge degeneracy
point.~\cite{wallraff:04} For two qubits which are not directly coupled to each other [$J_{\perp}=J_z=0$ in
Eq.~(\ref{eq:gen-ham})], but both of which are coupled to the cavity mode with $g_1 = g_2 = g \neq 0$, the
cavity-qubit coupling can be integrated out to leading order by a Schrieffer-Wolff transformation $U$. The
resulting effective Hamiltonian is~\cite{imamoglu:99,blais:04}
\begin{eqnarray}
\hat{H}^\prime & = & \hbar \left (\omega + \frac{2 g^2}{\Delta_1}\hat{s}_{1,z}  +  \frac{2
g^2}{\Delta_2}\hat{s}_{2,z} \right) \hat{a}^\dagger \hat{a} \label{eq:2qubit0}
\\ && + \hbar  \left(\Omega_1 + \frac{g^2}{\Delta_1}\right) \hat{s}_{1,z}
+  \hbar  \left(\Omega_2 + \frac{g^2}{\Delta_2}\right)
\hat{s}_{2,z}\nonumber \\
&& + \frac{\hbar}{2} \left(\frac{g^2}{\Delta_1} +
\frac{g^2}{\Delta_2} \right) \left(\hat{s}_1^+  \hat{s}_2^- +
\hat{s}_1^- \hat{s}_2^+  \right), \nonumber
\end{eqnarray}
with $\Delta_{1,2} = \Omega_{1,2}-\omega$.
Equation~(\ref{eq:2qubit0}) implies that (i) photon emission of one
qubit into the cavity and subsequent absorption by the other qubit gives rise
to an effective cavity-mediated inter-qubit coupling; and (ii) the cavity
resonance is energetically shifted relative to $\omega$ by
$g^2(-\Delta_1^{-1} - \Delta_2^{-1})$, $g^2(\Delta_1^{-1} -
\Delta_2^{-1})$, $g^2(-\Delta_1^{-1} + \Delta_2^{-1})$, and
$g^2(\Delta_1^{-1} + \Delta_2^{-1})$ for the two-qubit states
$|\!\downarrow \downarrow \rangle$, $|\! \uparrow \downarrow \rangle$,
$|\! \downarrow \uparrow  \rangle$, and $|\! \uparrow \uparrow \rangle$,
respectively. In particular, for $|\Delta_1 - \Delta_2|
\gg g^2|(\Delta_1^{-1} + \Delta_2^{-1})|$, spin flip-flop
transitions between the qubits are weak and the cavity resonance
can be used for the simultaneous readout of both qubits if the
cavity loss rate is sufficiently small that all four frequencies
can be resolved.~\cite{blais:04}

\subsection{Identical qubits with direct coupling}
\label{sec:disp-sw}

We now turn to the 
limit of identical qubits ($\Omega_1 = \Omega_2 = \Omega$ and $g_1=g_2=g$) with
finite direct coupling. Because
\begin{equation}
\hat{N} = \hat{a}^\dagger \hat{a} + \hat{s}_{1,z} + \hat{s}_{2,z}
\label{eq:Ndef}
\end{equation}
commutes with $\hat{H}$, the problem can be solved exactly by numerical
diagonalization of the Hamiltonian in the four-dimensional
subspaces with given eigenvalue $N$ of $\hat{N}$ (see below). First, we provide a perturbative
analytical solution in the limit of weak coupling where the
detuning of $\omega$ relative to all transitions of the coupled
qubit system is large compared to $g \sqrt{N}$. For $g=0$,
$\hat{H}$ is diagonal in the singlet - triplet basis, $|S\rangle =
(|\!\uparrow \downarrow\rangle - |\!\downarrow \uparrow
\rangle)/\sqrt{2}$, $|T_-\rangle = |\!\downarrow \downarrow\rangle$,
$|T_0\rangle = (|\!\uparrow \downarrow\rangle + |\!\downarrow
\uparrow \rangle)/\sqrt{2}$, and $|T_+\rangle = |\!\uparrow
\uparrow\rangle$ with energies $E_S = -  J_z/4 - J_\perp/2$, $E_-
= - \Omega+ J_z/4$, $E_0 = -  J_z/4 + J_\perp/2$, and $E_+=
\Omega+ J_z/4$. We now generalize the Schrieffer-Wolff
transformation of Refs.~\onlinecite{blais:04,imamoglu:99} for
finite $J_\perp$ and $J_z$ by defining
\begin{equation}
\hat{H}^\prime = U \hat{H} U^\dagger \label{eq:swtrafo1},
\end{equation}
where
\begin{eqnarray}
U & = & e^{\hat A} \label{eq:swtrafo2}, \\
\hat{A} & = & \frac{g}{2 \Delta_-}\left\{ \hat{a}\left[
\hat{s}_1^+ (1 - \hat{s}_{2,z} ) +  \hat{s}_2^+ (1 - \hat{s}_{1,z}
) \right] -{\rm H.c.} \right\} \nonumber \\
&&  + \frac{g}{2\Delta_+}\left\{ \hat{a}\left[ \hat{s}_1^+ (1 +
\hat{s}_{2,z} ) +  \hat{s}_2^+ (1 + \hat{s}_{1,z} ) \right] -{\rm
H.c.} \right\}, \nonumber
\end{eqnarray}
with $\Delta = \Omega - \omega$, $J = (J_{\perp} - J_z)/2$, and where
$\Delta_{\pm}=\Delta \mp J$ denotes the energy difference between
the triplet state $|T_{\pm}\rangle$ and $|T_0\rangle$. By
construction, the Schrieffer-Wolff transformation removes the
cavity coupling to leading order. Expanding $\hat{H}^\prime$ to
second order in $g$, we find
\begin{widetext}
\begin{eqnarray}
\hat{H}^\prime & = & \hbar \left[\Omega (\hat{n} + \hat{s}_{1,z} +
\hat{s}_{2,z}) -  \Delta \hat{n} + \frac{ J_\perp}{2}
\left( \hat{s}_1^+  \hat{s}_2^- + \hat{s}_1^- \hat{s}_2^+ \right)
+ J_z \hat{s}_{1,z} \hat{s}_{2,z} \right. \label{eq:gen-ham-transf}\\
& & + g^2 \left(\frac{\hat{n}+1}{\Delta_+} +
\frac{\hat{n}}{\Delta_-} \right) (\hat{s}_{1,z} + \hat{s}_{2,z})
+ 2 g^2 (2 \hat{n}+1) \left(\frac{1}{\Delta_+} -
\frac{1}{\Delta_-} \right) \hat{s}_{1,z}  \hat{s}_{2,z} + g^2
\left(\frac{\hat{n}+1}{\Delta_-} - \frac{\hat{n}}{\Delta_+}
\right) \left( \hat{s}_1^+  \hat{s}_2^- + \hat{s}_1^-
\hat{s}_2^+ \right) \nonumber \\
&& \left. + g^2 \left(\frac{1}{\Delta_+} - \frac{1}{\Delta_-} \right)
\left(\hat{a}^2 \hat{s}_1^+ \hat{s}_2^+ + \hat{a}^{\dagger 2}
\hat{s}_1^- \hat{s}_2^- \right)\right], \nonumber
\end{eqnarray}
\end{widetext}
where $\hat{n} = \hat{a}^\dagger \hat{a}$ is the photon number
operator. The second line of Eq.~(\ref{eq:gen-ham-transf}) is
diagonal in the singlet-triplet basis and can be interpreted as ac
Stark shift and Lamb shift of the qubit states due to the cavity mode. The third
line of Eq.~(\ref{eq:gen-ham-transf}) describes two-photon
transitions between $|T_+\rangle$ and $|T_-\rangle$.

The expansion in $g$ leading to Eq.~(\ref{eq:gen-ham-transf}) is
valid as long as $g \sqrt{N} \ll |\Delta_{\pm}|$. Unless 
$g^2N|\Delta_+^{-1} - \Delta_-^{-1}| \gtrsim |2\Delta + g^2N(\Delta_+^{-1} + \Delta_-^{-1})|$, 
two-photon
processes described by the last term in
Eq.~(\ref{eq:gen-ham-transf}) can be neglected. For a cavity mode
detuning which is off-resonant with all bare transition energies,
Eq.~(\ref{eq:gen-ham-transf}) implies that the cavity resonance
experiences a shift
\begin{equation}
\omega \rightarrow \omega + 2 g^2 \times \left\{
\begin{array}{cl}
0 & \hspace*{0.4cm} \textrm{for } |S\rangle,\\
\Delta_+^{-1} &  \hspace*{0.4cm} \textrm{for } |T_+\rangle, \\
\Delta_-^{-1}- \Delta_+^{-1} &  \hspace*{0.4cm} \textrm{for }
|T_0\rangle, \,\, \textrm{and} \\
-\Delta_-^{-1} &  \hspace*{0.4cm} \textrm{for } |T_-\rangle.
\end{array}
 \right. \label{eq:res-shift}
\end{equation}

Similarly to two qubits with different level spacing $\Omega_1
\neq \Omega_2$ and without direct coupling, $J_\perp = J_z = 0$, see Eq.~(\ref{eq:2qubit0}),
the cavity resonance splits into four lines, depending on the
state of the coupled qubits. In stark contrast to
qubits without a direct interaction, however, measurement of the
cavity resonance projects the
qubits onto the singlet-triplet basis rather than the
 product basis of $\hat{s}_{z}$ eigenstates. A resonance shift by $2
g^2(\Delta_-^{-1}- \Delta_+^{-1})$, corresponding to the
$|T_0\rangle$ state, indicates that the coupled qubits are in
a maximally entangled (and also super-radiant\cite{dicke:54}) state.
In contrast, the singlet state $|S \rangle$ (being a sub-radiant state) decouples from the radiation field due to symmetry reasons and does not induce a shift of the cavity resonance line.
However, such a proof of entanglement
would be rather indirect because it relies only on the energy
levels of the system rather than quantum state tomography.

\begin{figure}
\centerline{\mbox{\includegraphics[width=6.3cm]{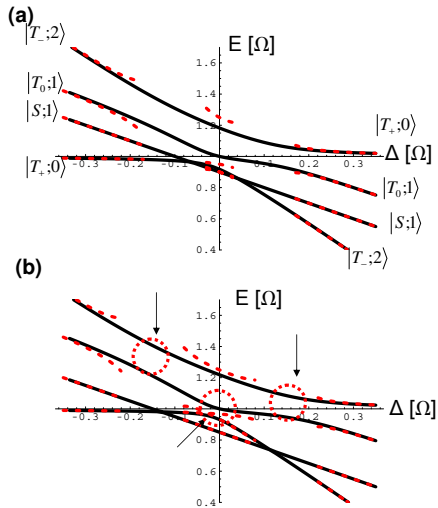}}}
\caption{(color online). Energy level spectrum in the subspace
$N=1$ as a function of $\Delta$ for $g=0.05 \Omega$ and (a) $J_\perp=0.2
\Omega$, $J_z=0$, and (b) $J_\perp=0.3 \Omega$, $J_z=0$, respectively. Solid lines indicate
the exact energy eigenvalues, dashed lines the approximate
analytical values obtained with Eq.~(\ref{eq:gen-ham-transf}). The
three anticrossings discussed in the text are indicated by arrows and circles
in (b).}\label{Fig1}
\end{figure}

In order to verify the range of validity of our results, we
compare the spectrum determined from Eq.~(\ref{eq:gen-ham-transf})
in the dispersive regime with exact diagonalization. In
Fig.~\ref{Fig1}, we show the exact level spectrum (solid lines) as
a function of $\Delta$ for $g=0.05 \Omega$, $J_z=0$ and (a)
$J_{\perp}=0.2 \Omega$ and (b) $J_{\perp}=0.3 \Omega$,
respectively. The analytical results valid in the perturbative
regime of large detuning from all resonances are indicated as
dashed lines.

\begin{figure}
\centerline{\mbox{\includegraphics[width=8cm]{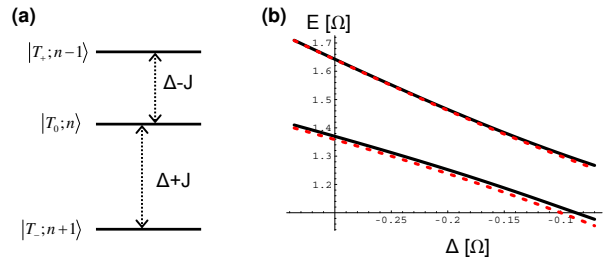}}}
\caption{(color online). (a) Triplet level scheme in the "dressed state" picture. (b) Magnification of the exact
anticrossing (solid lines) of $|T_-;2\rangle$ and $|T_0;1\rangle$ shown in Fig.~\ref{Fig1} (b), in comparison with a two-level
description (dashed lines).}\label{Fig2}
\end{figure}

As is evident from Fig.~\ref{Fig1}, the expansion in $g$ breaks down if the cavity mode approaches resonance
with one of the triplet transitions. On varying $\Delta$, the spectrum shows three anticrossings. Two strong
anticrossings with splitting $g \sqrt{2(n+1)}$ ($g \sqrt{2n}$) for $\Delta + J =0$ ($\Delta - J = 0$),
respectively, corresponding to strong mixing of the states $|T_-;n+1\rangle$ and $|T_0;n\rangle$
($|T_0;n\rangle$ and $|T_+;n-1\rangle$) by one-photon absorption. In addition, there is a weaker anticrossing
with splitting $2g^2 \sqrt{n(n+1)}/J$ between $|T_-;n+1\rangle$ and $|T_+;n-1\rangle$ at $\Delta =0$ due to
two-photon absorption processes described by the last term in Eq.~(\ref{eq:gen-ham-transf}). In Fig.~\ref{Fig2},
we show all relevant transitions and the exact energy eigenvalues in comparison with approximate expressions of
a two-state model which are obtained from a diagonalization in a truncated basis, taking only the two nearly-degenerate
states into account.
\begin{figure}[t]
\centerline{\mbox{\includegraphics[width=8.5cm]{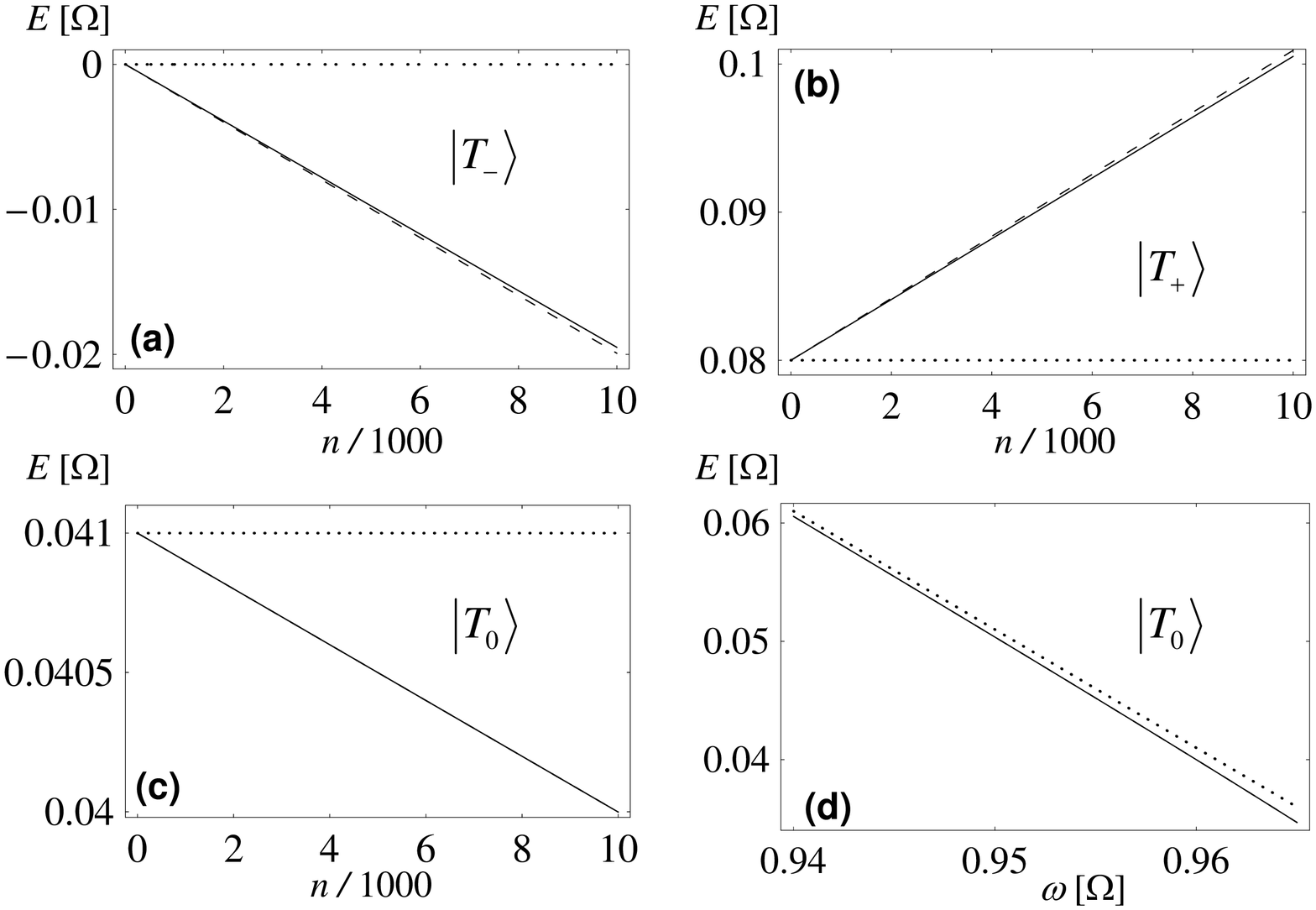}}}
\caption{Energies of two coupled identical qubits according to Eq.~(\ref{eq:gen-ham-transf}) (solid lines) and for the optical interaction without the RWA (dashed lines), see 
Eq.~(\ref{eq:starkfull}), of the triplet states (a) $|T_-\rangle$, (b) $|T_+\rangle$, and (c,d) $|T_0\rangle$ as a function of the photon number $n$ or the frequency $\omega$ of the cavity mode.  
The dotted lines indicate the energy levels in the absence of the coupling to the cavity, respectively.
The parameters used are $J=0.001\Omega$ and $g=0.0002\Omega$. Further,  $\omega = 0.96\Omega$ in (a,b,c), and $n=10^4$ in (d). }\label{FigStark}
\end{figure}

In Fig.~\ref{FigStark} we illustrate the ac Stark shift as described by the second line of Eq.~(\ref{eq:gen-ham-transf}) as a function of the cavity photon number $n$ and the frequency $\omega$. As two-photon transitions are negligible, the energy eigenvalues only depend on the average photon number $n$. We note that for large photon numbers in Fig.~\ref{FigStark}, $g\sqrt{n}\approx \Delta /2$ is reached, for which the perturbative coupling to the cavity is not rigorously satisfied.  We also show the ac Stark shift obtained for the optical interaction without the rotating-wave approximation (RWA), which we derive in Appendix \ref{sec:sw-general2}.

\subsection{Cavity-mediated two-qubit gate}

Before proceeding to the more general case of non-identical qubits coupled to a cavity, we discuss the effect of the direct qubit coupling on the proposed usage of the cavity as a quantum bus to couple the two qubits.\cite{blais:04,imamoglu:99} As shown in Ref.~\onlinecite{blais:04}, a $\sqrt{i\mathrm{SWAP}}$ two-qubit gate can be realized for $J_z = J_\perp =0$, when the two qubits are tuned into resonance with each other (while maintaining off-resonant coupling with the cavity) for the interaction time 
$t_{\mathrm{int}}=\pi \Delta / 4g^2$. 
However, in the presence of the direct qubit interaction, the two-qubit dynamics is determined by Eq.~(\ref{eq:gen-ham-transf}) rather than Eq.~(\ref{eq:2qubit0}), and the effective interaction via the cavity is accompanied by the direct qubit coupling. For simplicity, we focus on  the case $J_z = 0$ here,  which is realized, e.g., for capacitively coupled Cooper pair boxes, see Sec.~\ref{sec:mic-cpbs}.
The total effective two-qubit coupling energy in Eq.~(\ref{eq:gen-ham-transf}) is given by 
\begin{eqnarray}
\hbar \zeta \equiv \langle \uparrow \downarrow| \hat{H}' |\!\downarrow \uparrow \rangle& = & \frac{\hbar J_\perp}{2} + \hbar g^2 \left(\frac{\hat{n}+1}{\Delta_-} -\frac{\hat{n}}{\Delta_+}\right)\label{eq:eff2qubitcoup}\\
 & \simeq &  \frac{\hbar g^2}{\Delta} + \frac{\hbar J_\perp}{2}\left[1 - \frac{g^2}{\Delta^2} (2\hat{n}+1) \right]  ,\nonumber
\end{eqnarray}
where the second line is valid for $J_\perp \ll 2\Delta$.  
We quantify the fidelity of the $\sqrt{i\mathrm{SWAP}}$ gate proposed in Ref.~\onlinecite{blais:04} as $ f(t)=  \langle \uparrow\downarrow | U^\dagger_{\!\sqrt{i\mathrm{SWAP}}}\,U(t)| \!\uparrow\downarrow\rangle $, where $U_{\!\sqrt{i\mathrm{SWAP}}}$ is the ideal gate operator and $U(t)$ is the time-evolution operator for the  two qubits due to Eq.~(\ref{eq:gen-ham-transf}). Using the first line of Eq.~(\ref{eq:eff2qubitcoup}), we obtain for the gate fidelity
\begin{eqnarray}
f(t_{\mathrm{int}}) & = & \cos{\left(\zeta t_{\mathrm{int}} - \pi /4\right)}  \label{eq:gatefidelity} \\
 & \simeq & 1 - \frac{\pi^2 J_\perp^2 \Delta^2}{128 g^4} \left[1 - \frac{g^2}{\Delta^2} (2\hat{n}+1)\right]^2. \nonumber
\end{eqnarray}
Here, we also give the asymptotics for  
$J_\perp \ll 2 g^2/ \Delta$, when the cavity-mediated interqubit coupling dominates over the direct exchange-like coupling quantified by $J_\perp$.
We notice that $f(t_{\mathrm{int}})\approx 0$ for $J_\perp \approx 16 g^2/ \Delta$.  
For example, for a coupling $\hbar J_\perp = 0.1\,\mathrm{neV}$ ($\hbar J_\perp = 0.2\,\mathrm{neV}$) and the parameters\cite{schuster:05} $\hbar g \approx 0.01\,\mu\mathrm{eV}$, $\hbar \Delta \approx 1\,\mu\mathrm{eV}$, we obtain $f(t_{\mathrm{int}}) \approx 0.924$ $[f(t_{\mathrm{int}}) \approx 0.708]$ for a photon number $n=10$, indicating that a residual direct inter-qubit coupling may lead to gate errors in cavity-mediated two-qubit gates. However, this error may be significantly reduced by 
adjusting $t_{\mathrm{int}}$, as discussed in the following.

In order to reduce the fidelity loss discussed above, a two-qubit gate could be implemented based on the {\it total} two-qubit coupling energy Eq.~(\ref{eq:eff2qubitcoup}), in particular for systems for which the qubit interaction via the direct coupling and via the cavity are of comparable strength.
If both qubit energies are strongly detuned from $\omega$ and, additionally,  
$|\Omega_1 -\Omega_2| \gg J_\perp$, 
the direct qubit interaction is strongly suppressed along with the effective interaction via the cavity. Starting in such a situation, a $\sqrt{i\mathrm{SWAP}}$ two-qubit gate could be realized with  Eq.~(\ref{eq:eff2qubitcoup}) by tuning the two qubits into resonance and establishing simultaneously a smaller detuning of the qubits to the cavity. For the total two-qubit interaction, we readily obtain the required interaction time for a $\sqrt{i\mathrm{SWAP}}$ gate (up to phase factors) along the lines of Ref.~\onlinecite{blais:04} as 
$t_{\mathrm{int}}' = \pi/4 \zeta$. 
If a non-zero coupling $J_z$ is introduced, the block-diagonal form of $U(t)$ is preserved, and the $2\times 2$ subspace spanned by $|\!\uparrow\downarrow\rangle$ and $|\!\downarrow\uparrow\rangle$ contains a time-dependent phase factor $\exp{(-iJ_zt/2)}$ with respect to the $2\times 2$ subspace spanned by $|\!\uparrow\uparrow\rangle$ and $|\!\downarrow\downarrow\rangle$. 
We note that for the special case
$J_z /4 = J_\perp /2 + g^2[(n+1)/\Delta_- - n/\Delta_+]$, with a fixed photon number $n$, the total coupling of the two qubits is of the Heisenberg form, enabling a $\sqrt{\mathrm{SWAP}}$ two-qubit operation rather than $\sqrt{i\mathrm{SWAP}}$.

\subsection{Non-identical qubits with direct coupling}
\label{sec:sw-general}

Here, we provide the Schrieffer-Wolff transformation for the general case of arbitrary system parameters
$\Omega_1$, $\Omega_2$, $g_1$, and $g_2$ in the Hamiltonian Eq.~(\ref{eq:gen-ham}). For vanishing qubit-cavity
coupling, $g_1=g_2=0$, the eigenstates of the coupled qubit system are
\begin{eqnarray}
|T_+\rangle &=& |\! \uparrow \uparrow \rangle, \nonumber \\
|T_-\rangle &=& |\! \downarrow \downarrow  \rangle, \label{eq:gen-es} \\
|\psi_{s}\rangle & = & \cos(\phi/2) |\! \uparrow \downarrow \rangle +  \sin(\phi/2) |\! \downarrow \uparrow
\rangle, \hspace*{0.2cm} \textrm {and} \nonumber \\
|\psi_{a}\rangle & = & \sin(\phi/2) |\! \uparrow \downarrow \rangle -  \cos(\phi/2) |\! \downarrow \uparrow
\rangle \nonumber
\end{eqnarray}
with eigenenergies
\begin{eqnarray}
\hbar E_{\pm} &=& \hbar \frac{J_z}{4} \pm \hbar  \frac{\Omega_1 + \Omega_2}{2}, \label{eq:gen-eenergies} \\
\hbar E_{s/a}&=& -\hbar  \frac{J_z}{4} \pm \hbar  \frac{1}{2} \sqrt{(\Omega_1 - \Omega_2)^2 + J_\perp^2}
\nonumber,
\end{eqnarray}
respectively, where $\tan \phi = J_\perp/(\Omega_1 - \Omega_2)$.

\begin{figure}
\centerline{\mbox{\includegraphics[scale=0.35]{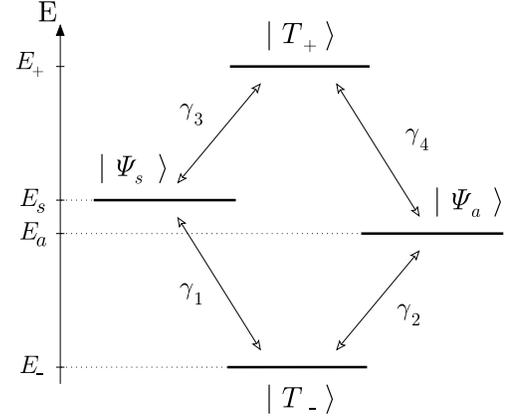}}} \caption{Energy level scheme of two coupled qubits
for the general case. The arrows indicate transitions under photon absorption and emission. Contrary to the
symmetric case discussed in Sec.~\ref{sec:dispersive}, for the general case there are four allowed transitions
with coupling constants $\gamma_i$, $i=1 \ldots 4$.}\label{Fig4}
\end{figure}

In contrast to the symmetric case discussed in Sec.~\ref{sec:dispersive},
where the singlet state decouples completely from the cavity field and only two transitions are 
allowed by
selection rules, there are four allowed transitions $|T_-\rangle \leftrightarrow |\psi_{s/a}\rangle$ and
$|\psi_{s/a}\rangle  \leftrightarrow |T_+\rangle$ in the general case, see Fig.~\ref{Fig4}. 
If the cavity mode
is off resonance with all four transitions, i.e., $|\omega - (E_+-E_{s/a})|, |\omega - (E_{s/a}-E_-)| \gg g_1
\sqrt{n}, g_2 \sqrt{n}$ with $n$ the characteristic photon number in the cavity, the cavity mode can be treated
perturbatively. In the eigenbasis Eq.~(\ref{eq:gen-es}), the Hamiltonian
Eq.~(\ref{eq:gen-ham}) reads
\begin{eqnarray}
\hat{H} &= &\sum_{i=\pm} \hbar E_i |T_i\rangle \langle T_i| +  \sum_{i=s,a} \hbar  E_i |\psi_i\rangle \langle
\psi_i| + \hbar \omega \hat{a}^\dagger \hat{a} \label{eq:gen-ham2} \\
&& + \hat{a} \hbar \left(\gamma_1 |\psi_s\rangle \langle T_-| + \gamma_2  |\psi_a\rangle \langle T_-| + \gamma_3
|T_+\rangle \langle \psi_s| \right. \nonumber \\ && \left. + \gamma_4 |T_+\rangle \langle \psi_a| \right) +
 \hat{a}^\dagger \hbar \left(\gamma_1 |T_- \rangle \langle \psi_s| + \gamma_2  | T_- \rangle\langle \psi_a |
 \right. \nonumber \\ && \left. + \gamma_3
|\psi_s \rangle \langle T_+ |  + \gamma_4 | \psi_a \rangle \langle T_+| \right) \nonumber,
\end{eqnarray}
with the coupling constants
\begin{eqnarray}
\gamma_{1,4} & = & \pm \cos(\phi/2) g_1 + \sin(\phi/2) g_2, \label{eq:gen-coupling-const} \\
\gamma_{2,3} & = & \sin(\phi/2) g_1 \mp  \cos(\phi/2) g_2.\label{eq:gen-coupling-const2}
\end{eqnarray}

To first order in $\gamma_i$, the Hamiltonian is diagonalized by the Schrieffer-Wolff transformation
\begin{equation}
\hat{H}^\prime = e^{\hat{a}\hat{B} - \hat{a}^\dagger \hat{B}^\dagger } \hat{H} e^{-\hat{a}\hat{B} +
\hat{a}^\dagger \hat{B}^\dagger } \label{eq:swtrafo3}
\end{equation}
with
\begin{eqnarray}
\hat{B} &=& \tilde{\gamma}_1 |\psi_s\rangle \langle T_-| +  \tilde{\gamma}_2 |\psi_a\rangle \langle T_-| \label{eq:swtrafo4} \\
&& +  \tilde{\gamma}_3 |T_+\rangle \langle \psi_s| +\tilde{\gamma}_4 |T_+\rangle \langle \psi_a|. \nonumber
\end{eqnarray}
The coefficients $\tilde{\gamma}_i$ are defined by
\begin{eqnarray}
\tilde{\gamma}_1 & = & -\frac{\gamma_1}{\omega - (E_s - E_-)} \equiv \frac{\gamma_1}{\Delta_{s-}},\nonumber \\
\tilde{\gamma}_2 & = & -\frac{\gamma_2}{\omega - (E_a - E_-)} \equiv \frac{\gamma_2}{\Delta_{a-}}, \label{eq:swtrafo5} \\
\tilde{\gamma}_3 & = & -\frac{\gamma_3}{\omega - (E_+ - E_s)} \equiv \frac{\gamma_3}{\Delta_{+s}}, \nonumber \\
\tilde{\gamma}_4 & = & -\frac{\gamma_4}{\omega - (E_+ - E_a)} \equiv \frac{\gamma_4}{\Delta_{+a}}. \nonumber
\end{eqnarray}
By assumption, $|\tilde{\gamma}_i| \ll 1$ because of the off-resonance condition. The expression for
$\hat{H}^\prime$ can then be truncated, neglecting terms of third and higher order in $\tilde{\gamma}_i$. Up to
second order in the coupling coefficients $\gamma_i$, we obtain
\begin{equation}
\hat{H}^\prime = \hat{H}_0 + \hat{H}_{\rm S,L} + \hat{H}_{\rm 2-ph} +  \hat{H}_{\rm mix}.
\label{eq:gen-ham-transf2}
\end{equation}
Here,
\begin{equation}
\hat{H}_0 = \sum_{i=\pm} \hbar E_i |T_i\rangle \langle T_i| +  \sum_{i=s,a} \hbar  E_i |\psi_i\rangle \langle
\psi_i| + \hbar \omega \hat{a}^\dagger \hat{a} \label{eq:h0}
\end{equation}
is the unperturbed Hamiltonian for vanishing cavity coupling.
\begin{eqnarray}
\hat{H}_{\rm S,L} & = & \hbar  \left(\hat{a}^\dagger \hat{a} + 1  \right)  \left(\tilde{\gamma}_3 \gamma_3 +
\tilde{\gamma}_4 \gamma_4  \right) |T_+\rangle\langle T_+| \nonumber \\
&& - \hbar  \hat{a}^\dagger \hat{a}  \left(\tilde{\gamma}_1 \gamma_1 + \tilde{\gamma}_2 \gamma_2  \right)
|T_-\rangle\langle T_-| \label{eq:gen-ham-transf2a} \\
&& + \hbar  \left[ \hat{a}^\dagger \hat{a} \left(\tilde{\gamma}_1 \gamma_1 - \tilde{\gamma}_3 \gamma_3  \right)
+
\tilde{\gamma}_1 \gamma_1 \right] |\psi_s\rangle \langle \psi_s|  \nonumber \\
&& + \hbar \left[ \hat{a}^\dagger \hat{a} \left(\tilde{\gamma}_2 \gamma_2 - \tilde{\gamma}_4 \gamma_4  \right) +
\tilde{\gamma}_2 \gamma_2 \right] |\psi_a\rangle \langle \psi_a| \nonumber
\end{eqnarray}
denotes the Stark and Lamb shift of the four basis states.
\begin{eqnarray}
\hat{H}_{\rm 2-ph} & = & \hbar  \frac{\tilde{\gamma}_3 \gamma_1 - \tilde{\gamma}_1 \gamma_3 + \tilde{\gamma}_4
\gamma_2
- \tilde{\gamma}_2 \gamma_4  }{2} \label{eq:gen-ham-transf2b} \\
&& \times \left(\hat{a}^2 |T_+\rangle \langle T_-| + \hat{a}^{\dagger 2} |T_-\rangle \langle T_+| \right)
\nonumber
\end{eqnarray}
describes two-photon transitions between $|T_+\rangle$ and $|T_-\rangle$.
\begin{eqnarray}
\hat{H}_{\rm mix} & = & \hbar  \frac{(\hat{a}^\dagger \hat{a} +1)\left( \tilde{\gamma}_1 \gamma_2 +
\tilde{\gamma}_2 \gamma_1 \right) -  \hat{a}^\dagger \hat{a}\left( \tilde{\gamma}_3 \gamma_4 + \tilde{\gamma}_4
\gamma_3 \right)}{2} \nonumber \\ && \times  \left( |\psi_s\rangle \langle \psi_a| + |\psi_a\rangle \langle
\psi_s| \right) \label{eq:gen-ham-transf2c}
\end{eqnarray}
is a term which gives rise to {\it transitions} between the basis states $|\psi_s\rangle$ and
$|\psi_a\rangle$. Its physical meaning will be explained in more detail in Appendix~\ref{sec:asym}. We note that for $J_\perp =0$
the effective interaction of $|\!\uparrow\downarrow\rangle$ and $|\!\downarrow\uparrow\rangle$ via the cavity
is given by $\hat{H}_{\rm mix}$. On a
technical level, $\hat{H}_{\rm S,L}$ and $\hat{H}_{\rm mix}$ correspond to virtual processes
corresponding to the emission and re-absorption of a photon. 

The cavity line shift induced by $\hat{H}_{\rm S,L}$ for the coupled-qubit eigenstates is given by 
\begin{equation}
\omega \rightarrow \omega +  \left\{
\begin{array}{cl}
 \gamma_3^2 / \Delta_{+s} + \gamma_4^2 / \Delta_{+a} & \hspace*{0.4cm} \textrm{for } |T_+\rangle,\\
-\gamma_1^2 / \Delta_{s-} - \gamma_2^2 / \Delta_{a-} &  \hspace*{0.4cm} \textrm{for } |T_-\rangle, \\
\gamma_1^2 / \Delta_{s-} - \gamma_3^2 / \Delta_{+s} &  \hspace*{0.4cm} \textrm{for } |\psi_{s}\rangle, \\
\gamma_2^2 / \Delta_{a-} - \gamma_4^2 / \Delta_{+a} &  \hspace*{0.4cm} \textrm{for } |\psi_a \rangle,
\end{array}
 \right. \label{eq:res-shift-gen}
\end{equation}
where the coupling coefficients $\gamma_i$ are given by Eqs.~(\ref{eq:gen-coupling-const}) and
(\ref{eq:gen-coupling-const2}), and the effective detunings $\Delta_{nm}$ are defined by Eq.~(\ref{eq:swtrafo5}). However, above states are only eigenstates of the qubits coupled to the cavity when $\hat{H}_{\rm mix} + \hat{H}_{\rm 2-ph}$ can be neglected. 
For $|J_z| \ll | J_\Omega | \equiv (1/2)|\sqrt{(\Omega_1 - \Omega_2)^2 + J_\perp^2}\,| \ll |\tilde \Delta | \equiv |(\Omega_1 + \Omega_2)/2 - \omega |$, the mixing energy of $|\psi_s\rangle$ and $|\psi_a\rangle$ due to $\hat{H}_{\rm mix}$ is approximately given by $\hbar [\sin\phi (g_1^2 -g_2^2)(2\hat{n}+1) - 2g_1g_2\cos\phi]/2 \tilde \Delta$. It is typically small due to the perturbative coupling condition and vanishes for coupled identical qubits ($\phi = \pi /2$ and $g_1=g_2$). 
Similarly, in the limit $|J_\Omega | \ll |\tilde \Delta |$, the mixing energy of $|T_+\rangle $ and $|T_- \rangle$ due to $\hat{H}_{\rm 2-ph}$ is proportional to $\hbar \sin\phi (g_1^2 +g_2^2)J_\Omega / \tilde \Delta^2$ and is therefore negligible.

The above analysis has been focussed onto coherent quantum dynamics. In real systems, however, the rate $\Gamma $ for spontaneous photon emission and the decay rate $\kappa $
of the cavity field lead to spectral line broadening.
For the Jaynes-Cummings model in the dispersive regime, the decay rates of the two dressed states of lowest energy (i.e., in the subspace of one energy quantum) are already given in Ref.~\onlinecite{blais:04}. Similar to the readout of
two qubits with no direct interaction,\cite{blais:04} we expect that the line shifts for directly coupled qubits can be well resolved if they are larger than $\kappa + \Gamma$. A more detailed treatment of
dissipation is beyond the scope of the present work.

\section{Control of the direct qubit interaction via photons}
\label{sec:disp-control}

As a second application of two qubits interacting with a cavity, in addition to the readout discussed above, we show that the qubits can be tuned into and
out of resonance by controlling the photon number in the cavity mode if the qubit-cavity couplings are not
identical. Here, we focus on the case of identical qubit level spacing, $\Omega_{1} = \Omega_{2}$ with strongly
asymmetric cavity couplings, $g_1 = g \neq 0$, $g_2 = 0$. As shown in Appendix \ref{sec:asym}, the Stark and
Lamb shift and mixing Hamiltonian, $H_{\mathrm{S,L}} + H_{\mathrm{mix}}$, then induce a relative energy shift of the states $|\! \uparrow\downarrow\rangle$  and $|\! \downarrow\uparrow \rangle$ which can be used to control the level admixing induced by the direct coupling
$J_\perp$.

For illustration, we focus on a cavity mode close to resonance with the transitions from $|T_+\rangle $ to $|\!\downarrow\uparrow \rangle$, $|\omega-(\Omega+J_z/2)| \ll |\omega-(\Omega-J_z/2)|$, 
but still in the dispersive
regime, $|g\sqrt{n}/[\omega-(\Omega+J_z/2)]| \ll 1$. Then, the ac Stark shift due to qubit $1$ is given by
\begin{eqnarray}
\hat{H}_{\rm S} & \simeq & \hbar \frac{g^2}{\omega-(\Omega+J_z/2)} \nonumber \\
 & & \times \left[ \hat{a}^\dagger \hat{a}
|\! \downarrow\uparrow \rangle \langle \downarrow\uparrow | - \left( \hat{a}^\dagger \hat{a} + 1\right)|T_+\rangle
\langle T_+| \right].\label{eq:starkasym}
\end{eqnarray}
If the number of photons in the cavity is $n$ such that
\begin{equation}
\left| \frac{g^2}{\omega-(\Omega+J_z/2)} \, n \right| \gg J_\perp,\label{eq:decouple}
\end{equation}
the two qubits are effectively decoupled due to the ac Stark shift, since level mixing due to $J_\perp$ is strongly suppressed. 
 Extracting the cavity photons on a
time scale short compared to $\tau_{\mathrm{sw}}=2\pi /J_\perp$ and maintaining the empty cavity state for a predetermined time ensures that, in the absence of an ac Stark shift, the direct interqubit coupling $J_\perp$ induces quantum coherent oscillations between $|\!\uparrow\downarrow\rangle $ and $|\!\downarrow\uparrow\rangle$. By reinserting a number of photons satisfying Eq.~(\ref{eq:decouple}) faster than the time $\tau_{\mathrm{sw}}$ into the cavity, the two qubits can be decoupled again.
Based on the latter inequality we define a minimum photon number for optical switching of the direct coupling, $n_{\mathrm{min}}= J_{\perp}|\omega -(\Omega +J_z/2)|/g^2$. 
However, in order for the analysis leading to Eq.~(\ref{eq:starkasym}) to remain valid, the perturbative coupling condition $n_{\mathrm{min}} g^2 \ll |\omega -(\Omega +J_z/2)|^2$ must be satisfied. From this follows $|\omega - (\Omega +J_z/2)|\gg J_{\perp}$ and, using the definition of $n_{\mathrm{min}}$, $n_{\mathrm{min}} \gg (J_\perp / g)^2$. 
Obviously, decoupling of the two qubits requires a larger photon number $ n >  n_{\mathrm{min}}$ in the case when {\it both}
coupling constants are non-zero (but still satisfy $|g_1| \neq |g_2|$). 
In Section \ref{sec:microscopic} we study three example systems for the coupled qubits: excitons in quantum dots, excitons in quantum shells, and capacitively coupled Cooper-pair boxes. For those systems, we calculate the parameters $J_\perp$ and $J_z$
for typical coupling mechanisms. Based on these estimates, we present in Table \ref{tab:switchparam} estimates for the switching time $\tau_{\mathrm{sw}}$ and the minimum
photon number $n_{\mathrm{min}}$ required to switch the direct
coupling for the energy scales of $\hbar J_\perp$ obtained in Section \ref{sec:microscopic}. 
In Table \ref{tab:switchparam}, we consider detunings satisfying $|\omega -(\Omega +J_z/2)|\geq J_{\perp}$ within the given range of $J_{\perp}$, respectively. For the maximum values of $J_{\perp}$,  these detunings and $n_{\mathrm{min}}$ do not strictly satisfy the condition for perturbative coupling to the cavity and thus provide a lower bound for $n_{\mathrm{min}}$.

While the addition of photons to the cavity mode can be realized via external irradiation, a fast extraction requires
a cavity design enabling a quick reduction of the Q-factor. For optical microcavities containing Bragg mirrors,
this could be achieved, e.g., via optical switching of the reflectivity.\cite{hastings:05} Such a scheme would
facilitate an all-optical control of the interaction of two qubits with possible applications as an optically controlled 
two-qubit gate. However, a high degree of control on the parameters of a high-Q cavity is required to perform
such a gate operation via a cavity mode.  In particular, the quick reduction of the Q-factor still needs to be
demonstrated experimentally.
As an interesting  alternative, an ac Stark shift for exciton qubits in quantum dots or shells can also be induced by a laser instead of a cavity. 
The switching "on" and
"off" of the direct qubit coupling might then be accomplished with ultrafast optical techniques. As a recent
first step into this direction, the laser-induced ac Stark effect has been demonstrated experimentally for
single GaAs quantum dots.\cite{unold:04}
\begin{table}[t]
\begin{center}
\begin{tabular}{|l|c|c|c|}
\hline
   & QD & QS & CPB\\
 \hline
 $\tau_{\mathrm{sw}}\; [\mathrm{ps}]$ & $0.4 \dots 4$ & $4 \dots 40$ & $400 \dots 4000$\\
\hline
$n_{\mathrm{min}}$ & $10^3 \dots 10^4$ & $10 \dots 10^2$ & $10^{5} \dots 10^{6}$\\
\hline
 $\hbar J_{\perp}\; [\mathrm{meV}]$ & $1 \dots 10$ & $0.1 \dots 1$ & $10^{-3} \dots 10^{-2}$\\
\hline
$\hbar g \; [\mathrm{meV}]$ & $ 0.1 $ & $ 0.1 $ & $10^{-5}$ \\
\hline
 $\hbar|\omega -(\Omega +J_z/2)|\; [\mathrm{meV}]$ & $10$ & $1$ & $0.01$ \\
\hline
\end{tabular}
\end{center}
\caption{Switching time $\tau_{\mathrm{sw}}=2\pi /J_\perp$ and lower bound (see text) for the minimum photon number $n_{\mathrm{min}}=J_\perp |\omega -(\Omega +J_z/2)|/g^2$ to decouple two qubits with $\Omega_1 = \Omega_2$ and $g_1 = g \neq g_2=0$ via the
ac Stark shift in a cavity. The considered types of qubits are excitons in coupled quantum dots
(QD), excitons in quantum shells (QS), and capacitively coupled Cooper pair boxes (CPB) with estimates for $J_\perp$ and $J_z$ from Sec.~\ref{sec:microscopic}. The estimates for $\hbar g$ are based on
Refs.~\onlinecite{schuster:05,badolato:05}.\label{tab:switchparam}}
\end{table}

Complementary to the decoupling of two identical qubits with asymmetric coupling to a cavity (or laser) mode,
the ac Stark shift could be exploited to bring two non-identical qubits into resonance, as already suggested by
Nazir {\it et al.}~\cite{nazir:04} for excitons in quantum dots coupled to a laser. 
The control of the direct interaction $J_\perp$ could then be achieved via
the shaping of microwave or laser pulses, for superconducting or exciton qubits, respectively.
We note that in our work the exact Schrieffer-Wolff transformation is provided for the coupled system with and without the RWA for the optical interaction, while Nazir {\it et al.} provided a numerical solution within the RWA.

The coupling constants $g_1$ and $g_2$ of the qubits to the cavity may be intrinsically different due to
fluctuations in the composition, size, and geometry of the structure containing the qubit, which may randomly
occur or might be induced on purpose in the fabrication process. In principle, asymmetric coupling of the two
qubits to the cavity could also be achieved due to the different qubit locations with respect to the cavity field. 
Still, the distance between coupled
quantum dots or superconducting islands is typically much smaller than the cavity-photon wavelength, which complicates this approach.  
A strongly
asymmetric cavity coupling could also be achieved via the misalignment of the electric dipole moment $\mathbf{d}_i$ of one qubit relative to
the electric field in the cavity, e.g., for Cooper-pair boxes in a line resonator. Ideally, (say) $\mathbf{d}_2$ is oriented perpendicular and $\mathbf{d}_1$ is oriented parallel to the
cavity field, such that the coupling of electric dipole transitions to the cavity is strongly asymmetric.

\section{Microscopic Models}
\label{sec:microscopic}

We next derive explicit expressions for the characteristic energy scales in Eq.~(\ref{eq:gen-ham}) for qubits
defined in terms of exciton states of a given spin in quantum dots or quantum shells and for superconducting qubits at the
charge degeneracy point. While most quantum computing schemes for quantum dots rely on the electron spin as
qubit,~\cite{cerletti:05a} we focus on charge-based qubits defined by the presence or absence of an exciton (see, e.g.,
Ref.~\onlinecite{biolatti:00}), because such states couple to the cavity mode by photon emission and absorption. Here, we rely on the fact that the exciton spin relaxation time is typically much longer than the exciton lifetime in quantum dots at low temperatures and low magnetic fields.~\cite{tsitsishvili:03} Optical selection rules then ensure the coupling of the polarized 
excitons to a cavity mode with a suitable polarization, see, e.g., Ref.~\onlinecite{cerletti:05b} for  typical optical transition matrix elements for quantum dots.
The integration of quantum dots into high-Q cavities is still a
considerable challenge, but has been demonstrated for self-assembled quantum dots~\cite{badolato:05} and also
appears to be feasible by deposition of chemically synthesized nanocrystals onto defect modes of photonic
crystals.~\cite{alivisatos:96} For definiteness, we focus on the latter system in the following.

The analysis can be easily extended to electron spin qubits in quantum dots, for which the electron exchange interaction leads to an effective coupling of the Heisenberg form [i.e., $J_\perp = J_z/2$ in Eq.~(\ref{eq:gen-ham})], provided that the single-dot charging energy is larger than the tunnel matrix element, which has already been studied in detail in Refs.~\onlinecite{loss:98,burkard:99,burkard:00}. For electron spins, a static magnetic field gives rise to Zeeman splittings $\hbar \Omega_i$, typically on the order of tens of $\mu$eV for fields on the order of 1 Tesla, and the two spins couple, e.g.,  via magnetic dipole transitions to a microwave cavity.
Alternatively, cavity-mediated spin interactions are implemented by two-photon Raman 
transitions via trion states.~\cite{imamoglu:99}  We note that the spatial separation of coupled dots is typically too small to enable the application of individual laser fields to the two dots. 
Still, effective spin interactions could be mediated via the cavity if the electrons in the two different dots have different Zeeman splittings, e.g., due to magnetic dopants\cite{imamoglu:00} (giving different gyromagnetic factors), magnetic field gradients, or
inhomogeneous hyperfine polarizations, since then two laser fields with different frequencies could be applied to the two dots simultaneously to establish the Raman transition conditions.
For the special case of identical dots, these conditions can
be satisfied for both dots even by the same laser field.

\subsection{Quantum dots}
\label{sec:mic-qd}

\begin{figure}
\centerline{\mbox{\includegraphics[width=8.5cm]{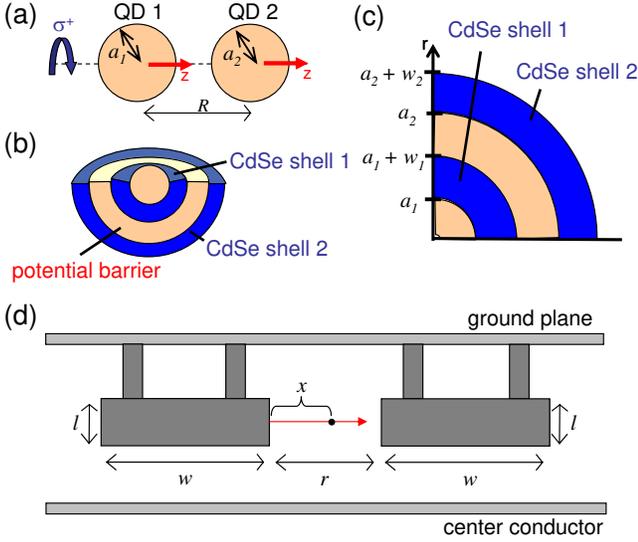}}}
\caption{(color online). Schemes for coupled qubits: (a) Pair of quantum dots interacting with one
cavity mode as discussed in Sec.~\ref{sec:mic-qd}. (b) Nested
quantum shells as discussed in Sec.~\ref{sec:mic-qs}. (c)
Cross-section through nanocrystal containing a pair of nested
quantum shells. (d) Two Cooper-pair boxes in a line resonator as discussed in Sec.~\ref{sec:mic-cpbs}.}\label{Fig3}
\end{figure}

The confinement of electrons in nanocrystal quantum dots gives rise to
well-defined quantum size levels with a spacing which is typically
of order $0.1~\,{\rm eV}$ for the conduction band. We first consider pairs of nanocrystals
interacting with a $\sigma^-$ circularly polarized cavity mode
[Fig.~\ref{Fig3}(a)], assuming that the symmetry axis of both
nanocrystals is aligned with the propagation direction of the cavity
photon. For definiteness, we will focus on materials
with valence band states described by the spherical Luttinger
Hamiltonian, which is a good approximation for, e.g., CdSe. The
cavity mode is assumed to be sufficiently close to the absorption
edge that only the lowest exciton transition from the $1S_{3/2}$
valence band multiplet to the conduction band ground state $1S_e$
must be taken into account. Crystal or shape anisotropy are
assumed to split the light-hole (lh) and heavy-hole (hh) states at
the $\Gamma$ point and only transitions involving hh states are
considered. The ground state of the nanocrystal, i.e., the exciton vacuum, is denoted as
$|\!\downarrow\rangle$, while the one-exciton state is denoted as
$|\!\uparrow\rangle$.

The transverse coupling term in Eq.~(\ref{eq:gen-ham}) corresponds
to coherent exciton transfer between the nanocrystals. Possible microscopic
mechanisms are direct transfer of electron and hole, e.g., via a
coupling molecule~\cite{ouyang:03,meier:04,schrier:05} or resonant F\"orster
transfer.~\cite{govorov:03,lovett:03,nazir:03} In order for
Eq.~(\ref{eq:gen-ham}) to capture the essential features of
molecularly coupled nanocrystals, indirect exciton states must be
negligible. For pairs of identical nanocrystals this implies that
the transfer matrix elements for both the conduction and valence
band, $t_c$ and $t_v$, respectively, must be small compared to the
exciton binding energy $U \simeq 1.6 e^2/4 \pi \epsilon_0 \epsilon
a$, with  the nanocrystal radius $a$. Data for CdSe nanocrystals
coupled by benzene rings~\cite{ouyang:03} are consistent with $t_c
\simeq 80 \, {\rm meV}$ while $U\simeq 160 \, {\rm meV}$ for $a=
1\, {\rm nm}$. Because $t_c$ and $U$ are of the same order of
magnitude, the transfer matrix element for an exciton,
$\hbar J_{\perp} = 2 t_c t_v/U \simeq t_v$, is determined by $t_v$. While
$t_v$ is expected to be smaller than $t_c$ because of the
larger hole mass, 
atomistic calculations for molecularly coupled nanocrystals have shown that electron and hole transmission probabilities are comparable.\cite{schrier:05}
The exciton transfer energy $\hbar J_{\perp}$ can thus presumably reach
values of order $10 \, {\rm meV}$.

Alternative mechanisms for exciton transfer are F\"orster
processes which result from the electrostatic dipole interaction
Hamiltonian~\cite{govorov:03,lovett:03,nazir:03}
\begin{equation}
\hbar J_{\perp}=\frac{e^2}{4 \pi \epsilon_0 \epsilon R^3} \left[\langle {\bf
r}_1\rangle \langle {\bf r}_2\rangle - \frac{3}{R^2} \left(\langle
{\bf r}_1\rangle \cdot {\bf R} \right) \left( \langle {\bf
r}_2\rangle \cdot {\bf R} \right)\right], \label{eq:el-dip}
\end{equation}
where ${\bf R}$ denotes the relative position of the nanocrystals
and $\langle {\bf r}_{1,2}\rangle$ are the dipole transition
matrix elements between the exciton vacuum and the one-exciton
states in nanocrystals $1$ and $2$, respectively. Because the
symmetry axes of both nanocrystals are aligned 
%with the propagation direction of the cavity photon 
by assumption [Fig.~\ref{Fig3}(a)],
the F\"orster transfer in Eq.~(\ref{eq:el-dip}) conserves the
exciton spin.~\cite{govorov:03,nazir:03} The dipole matrix element
connecting the $-3/2$ state of the $1S_{3/2}$ valence band
multiplet with the $s_z = -1/2$ conduction band ground state is
\begin{equation}
\langle 1S_e; \uparrow | {\bf r} | 1S_{3/2};-3/2 \rangle = \langle
S|x|X\rangle \frac{1}{\sqrt{2}} \left(\begin{array}{c} 1 \\ -i \\
0 \end{array} \right) I_{1S_{3/2};1S_e}. \label{eq:dipole}
\end{equation}
$\langle S|x|X\rangle$ denotes the interband matrix element. The
orbital overlap of the envelope wave functions of $1S_e$ and the
$L=0$ component of the  $1S_{3/2}$ multiplet,
\begin{equation}
I_{1S_{3/2};1S_e} = \int_0^a dr \, r^2 \psi_{1S_e}(r) R_0(r),
\end{equation}
is typically of order unity. For CdSe nanocrystals with a
high-bandgap capping layer,
$I_{1S_{3/2};1S_e}=0.88$.~\cite{efros:92} Inserting
Eq.~(\ref{eq:dipole}) into Eq.~(\ref{eq:el-dip}), we find for the
F\"orster transfer energy
\begin{equation}
\hbar J_{\perp} \simeq \frac{e^2 |\langle S|x|X\rangle|^2}{4 \pi
\epsilon_0 \epsilon R^3}.
\end{equation}
For CdSe, $|\langle S|x|X\rangle| \sim 1 \, {\rm nm}$, such that
$\hbar J_{\perp} = 1.44 \, {\rm eV}/\epsilon (R[{\rm nm}])^3$ is
of order $1 \, {\rm meV}$ for $R=5 \, {\rm nm}$ and $\epsilon=9$.

The longitudinal coupling term $\propto \hbar J_z $ in Eq.~(\ref{eq:gen-ham})
represents the biexciton shift. Dominant contributions are the
electrostatic quadrupolar interaction and van-der-Waals
interactions which correspond to simultaneous virtual transitions
between the $1S_{3/2}$ and $1P_{3/2}$ valence band multiplets in
both dots. The electrostatic interaction results from the charge
density of the $1S_{3/2}$ valence band
state,~\cite{xia:89,efros:92}
\begin{eqnarray}
&& |1S_{3/2};F_z=3/2\rangle  \label{eq:1sstate} \\ &&
\hspace*{1cm} = R_0(r) |L=0,j=3/2,F=3/2,F_z=3/2\rangle \nonumber \\
&& \hspace*{1.5cm}+ R_2(r) |L=2,j=3/2,F=3/2,F_z=3/2\rangle.
\nonumber
\end{eqnarray}
$R_{0,2}$ are the radial wave functions for an orbital angular
momentum $L=0,2$ of the envelope wave function.~\cite{xia:89,meier:04b} The electrostatic
quadrupole moment for a spherical quantum dot with radius $a$ and a
high-bandgap cap is
\begin{equation}
Q_{zz} = e \frac{4}{5} \int_0^a dr \,r^4 R_0 (r) R_2 (r)
\label{eq:quadrup1}
\end{equation}
for vanishing boundary conditions.
For a spherical CdSe nanocrystal with radius $a$, $Q_{zz} = - 0.06 e a^2$.
The biexciton shift resulting
from quadrupolar electrostatic interactions~\cite{jackson} is for $a \ll R$ given by
\begin{equation}
\hbar J_{z,Q} \simeq \frac{3Q_{zz}^2}{8\pi  \epsilon_0 R^5}  =  2.16 \,
{\rm eV} \frac{(Q_{zz}/e \,[{\rm nm^2}])^2}{(R[{\rm nm}])^5}.
\label{eq:quadrup2}
\end{equation}
Because the radial integral in Eq.~(\ref{eq:quadrup1}) is
typically small, the electrostatic contribution to the biexciton
shift is smaller than $1 \,{\rm meV}$ for characteristic
parameters. For CdSe nanocrystals, $a=1\, {\rm nm}$ ($2 \, {\rm
nm}$), and $R=5\, {\rm nm}$, $\hbar J_z = 2.4 \, \mu{\rm eV}$ ($39.8
\, \mu{\rm eV}$). However, Eq.~(\ref{eq:quadrup2}) is no longer rigorously applicable 
in this parameter range and may only provide an estimate for the characteristic energy scale.

For these or comparable parameters, the dominant contribution to the biexciton shift is caused by
inter-dot van-der-Waals interactions. Because the level spacing for
valence band states is small compared to the conduction band, the
dominant contribution involves simultaneous transitions between
the $1S_{3/2}$ and $1P_{3/2}$ valence band states in both quantum dots. In
second order perturbation theory in the hole-hole Coulomb interaction,
\begin{widetext}
\begin{equation}
 \hbar J_{z,vW} = \left(\frac{e^2}{4 \pi \epsilon_0 \epsilon R^3}
\right)^2 \sum_{F_{z,1},F_{z,2}} \frac{\left| \mbox{}_1\langle
1P_{3/2};F_{z,1}| \mbox{}_2 \langle 1P_{3/2};F_{z,2}| x_1 x_2 +
y_1 y_2- 2 z_1 z_2 |1S_{3/2};3/2\rangle_1 |1S_{3/2};3/2\rangle_2
\right|^2}{\Delta E_1 + \Delta E_2}. \label{eq:vw1}
\end{equation}
\end{widetext}
The matrix element is finite only for $F_{z,1}=F_{z,2}=3/2$, i.e.,
for processes which describe simultaneous transitions from
$|1S_{3/2};3/2\rangle$ to $|1P_{3/2};3/2\rangle$ in both dots.
$\Delta E_{1,2}$ denotes the energy difference between the
corresponding hole levels in quantum dot $1$ and $2$, respectively. For a pair
of identical dots,
\begin{equation}
 \hbar J_{z,vW} = \left(\frac{e^2}{4 \pi \epsilon_0 \epsilon R^3}
\right)^2 \frac{2 \left| \langle 1P_{3/2};3/2| z
|1S_{3/2};3/2\rangle \right|^4}{\Delta E}. \label{eq:vw2}
\end{equation}
The transition matrix element,
\begin{equation}
\langle 1P_{3/2};3/2| z|1S_{3/2};3/2\rangle \simeq \frac{1}{\sqrt{5}}
\int_0^a dr \, r^3 R_0(r) R_1(r) \label{eq:vw3}
\end{equation}
depends on the specific material parameters, but is typically of
order $0.1 a$. For CdSe, $\langle 1P_{3/2};3/2|
z|1S_{3/2};3/2\rangle = 0.17 a$, while $\Delta E = -27 \, {\rm meV}
/( a[{\rm nm}])^2$. For $a=1 \, {\rm nm}$ ($2 \, {\rm nm}$) and
$R=5 \, {\rm nm}$, the van-der-Waals contribution to the
biexciton shift is $\hbar J_{z,vW} = - 8.2 \, \mu{\rm eV}$ ($- 132 \,
\mu{\rm eV}$), larger than the electrostatic contribution derived
above.

We conclude that for nanocrystals with radius of order $1-2\, {\rm
nm}$ and a distance of order $5 \, {\rm nm}$, the energy scale for
F\"orster transfer, $1-10 \, {\rm meV}$, is large compared to the
inter-dot biexciton shift which is typically of order $\sim 10 \, \mu{\rm eV}$.

\subsection{Nested quantum shells}
\label{sec:mic-qs}

Similarly to quantum dots, quantum-dot quantum wells or 'quantum
shells'~\cite{battaglia:03,berezovsky:04,meier:04b} provide
quantum confinement for electrons and holes which allows one to
define a qubit by the presence or absence of an exciton with a given polarization. While
experimental studies have so far focussed on individual shells,
with a low-bandgap material sandwiched between a high-bandgap core
and cap, synthesis of nested shells is possible. Such a system is
expected to exhibit properties similar to a pair of quantum dots
[Figs.~\ref{Fig3}(b),(c)]. Here, we derive the characteristic
energy scales for exciton transfer between shells ($\hbar
J_\perp$) and the biexciton shift ($\hbar J_z$).
%The high symmetry of the system allows us to evaluate both the F\"orster coupling and the exciton shift without invoking the dipole approximation.
All energy scales typically depend on the inner and outer radius of a
shell, the potential barrier, and the band masses inside the shell
and the barrier. For definiteness, we consider shells with width
$w_i$ small compared to their inner radius $a_i$, where $i=1,2$,
in the limit of high potential barriers where both conduction and
valence band states are well confined inside the shells. Then, the
wave functions of the lowest conduction band state is
\begin{equation}
\psi_{i} ({\bf r}) = \sqrt{(2/w_i)} \, \frac{\sin[\pi(r-a_i)/w_i
]}{r} Y_{00}(\hat{\bf r}), \label{eq:shell-cbgs}
\end{equation}
while the wave function of the $1S_{3/2}$ valence band states with
total angular momentum quantum numbers $F=3/2$ and $F_z$ is
\begin{eqnarray}
 \phi_{i;F_z} &=& \sqrt{(1/w_i)} \, \frac{\sin[\pi(r-a_i)/w_i ]}{r}
\label{eq:shell-vbgs} \\ && \hspace*{0.5cm} \times
\left(|L=0;j=3/2;F=3/2;F_z\rangle \right. \nonumber \\
&& \hspace*{1cm}\left.  +|L=2;j=3/2;F=3/2;F_z\rangle \right).
\nonumber
\end{eqnarray}
The energy splitting to the $1P_{3/2}$ multiplet is $\hbar^2
(\gamma_1 - 2 \gamma)/ m_0 a_i^2$, where $m_0$ is the free
electron mass and $\gamma_1$ and $\gamma$ denote the Luttinger
parameters of the shell material.

Dominant mechanisms of exciton transfer between the shells are electron and hole tunneling or F\"orster
transfer. For a finite potential barrier $V_{0,i}$ of width $w$, the tunneling matrix elements for electrons
($i=c$) and holes ($i=v$) are exponentially suppressed with a factor
$\exp{(-2w\sqrt{2m_i(V_{0,i}-E_i)}/\hbar)}$. For realistic structures, such as CdSe/ZnS/CdSe nanocrystal
heterostructures, the electron tunneling matrix element can be of order $\sim 10-50 \, {\rm meV}$.\cite{meier:un} In order to
calculate the matrix element for the F\"orster process, we invoke the identity
\begin{equation}
\frac{1}{|{\bf r_1} - {\bf r_2}|} = \sum_{l;m} \frac{4\pi}{2l+1}
Y_{l-m}(\widehat{\bf r}_1) Y_{lm}(\widehat{\bf r}_2)
\frac{r_1^{l}}{r_2^{l+1}}, \label{eq:coul1}
\end{equation}
where $r_1< r_2$, i.e., the index $1$ denotes the inner shell.
%Angular momentum selection rules imply that the only finite matrix
%element for a F\"orster process between two concentric quantum shells
%comes from $l=1,m=\pm 1$ and $l=2, m= \pm 1$ in the expansion
%Eq.~(\ref{eq:coul1}).
Similarly to quantum dots, F\"orster transfer for nested
quantum shells conserves the exciton spin. In order to estimate the
characteristic energy scale of the F\"orster matrix element, we
calculate the transition matrix element for the dominant
$l=1$ -- term in Eq.~(\ref{eq:coul1}) and approximate the hole state by
its $L=0$ envelope function component. Then,
\begin{eqnarray}
&& \langle \downarrow_1;\uparrow_2 | \frac{1}{|{\bf r_1} - {\bf r_2}|} | \uparrow_1;\downarrow_2 \rangle \sim
\frac{4 \pi}{3} \frac{1}{3 \pi} \frac{2}{w_1 w_2}|\langle S|x|X\rangle|^2 \nonumber
\\ && \hspace*{1cm}\times    \int_{a_1}^{a_1+w_1} dr \sin^2 [\pi (r-a_1)/w_1] \,
  \label{eq:coul2}\\ &&  \hspace*{1cm} \times
\int_{a_2}^{a_2+w_2} dr \frac{\sin^2 [\pi (r-a_2)/w_2]}{r^3}
\nonumber \\
&  & \simeq  \frac{2 |\langle S|x|X\rangle|^2}{9 a_2^3}. \nonumber
\end{eqnarray}
The approximation in
%factor of $1/3\pi$ results from the integration over angular
%variables in both shells, while
the last line in
Eq.~(\ref{eq:coul2}) is valid in the limit of narrow shells
considered here. $\langle S|x|X\rangle$ denotes the Kane interband
matrix element. Assuming that the dielectric constant of shells
and potential barriers are comparable, Eq.~(\ref{eq:coul2}) implies
\begin{equation}
\hbar J_\perp \sim \frac{2 e^2 \, |\langle S|x|X\rangle|^2}{36 \pi \, \epsilon_0 \epsilon \, a_2^3},
\end{equation}
which only depends on the radius of the outer shell but is independent of the
size of the inner shell. For CdSe,
$\epsilon = 9$ and $|\langle S|x|X\rangle| \sim 1 \, {\rm nm}$,
such that $\hbar J_\perp \simeq 35 \, {\rm meV}/(a_2[{\rm
nm}])^3$. For small nanocrystals, where, e.g., $a_2 \simeq 4 \, {\rm nm}$, the F\"orster matrix element is of
order $\hbar J_\perp \simeq 0.5 \, {\rm meV}$.

Similarly to quantum dots, the dominant contributions to the biexciton
shift for quantum shells are the electrostatic quadrupolar
interaction caused by the non-spherical charge density of the
holes and van-der-Waals interactions. With the expansion
Eq.~(\ref{eq:coul1}), after
integration over the angular variables, we obtain for the
electrostatic quadrupolar hole interaction
\begin{eqnarray}
\hbar J_{z,Q} &=& \frac{e^2}{4\pi \epsilon \epsilon_0} \frac{4}{25}
\int_{a_1}^{a_1+w_1} dr_1 \, r_1^4 \, R_0(r_1) R_2(r_1)
\label{eq:shell-biexc} \\
&& \times \int_{a_2}^{a_2+w_2} dr_2 \, \frac{ R_0(r_2) R_2(r_2) }{r_2}
\nonumber \\ &=& \frac{e^2}{4\pi \epsilon \epsilon_0}
\frac{a_1^2}{25 a_2^3}, \nonumber
\end{eqnarray}
where, in the last line, the radial integrals have been evaluated
for the limit of narrow shells considered here,
Eq.~(\ref{eq:shell-vbgs}), and it is understood that the hole functions $R_{0,2}$ in Eq.~(\ref{eq:shell-biexc}) denote the corresponding components of the hole wave functions in shells 1 and 2, respectively. For $a_1=2 \, {\rm nm}$, $a_2=3 \, {\rm
nm}$, and $\epsilon = 9$, $\hbar J_{z,Q} = 0.9 \, {\rm meV}$.

The van-der-Waals interaction caused by simultaneous transitions
between $1S_{3/2}$ and $1P_{3/2}$ hole multiplets in both shells
can be calculated similarly to Eq.~(\ref{eq:vw1}) in second order
perturbation theory in the hole-hole Coulomb interaction. In terms
of the radial wave functions for the $1S_{3/2}$ and $1P_{3/2}$
valence band multiplets,
\begin{eqnarray}
\hbar J_{z,vW} &=& \left(\frac{e^2}{4\pi \epsilon_0 \epsilon}\right)^2
\left( \frac{4}{5}\right)^2 \frac{1}{\Delta E_1 + \Delta E_2}
\label{eq:vw4} \\
&& \times \left[\int_{a_1}^{a_1+w_1} \!\!\!\!\!\!\!\!\!\!\!\! dr_1 \, r_1^3  \left(\frac{R_0
R_1}{2} + \frac{2 R_2 R_1}{5} + \frac{3 R_2 R_3}{10}
\right)\right]^2 \nonumber \\
&& \times \left[\int_{a_2}^{a_2+w_2} \!\!\!\!\!\!\!\!\!\!\!\! dr_2  
\left(\frac{R_0 R_1}{2} + \frac{2 R_2 R_1}{5} + \frac{3 R_2
R_3}{10} \right)\right]^2 \nonumber \\
& \simeq & \left(\frac{e^2}{4\pi \epsilon_0 \epsilon} \, \frac{36
a_1}{125 a_2^2}\right)^2 \frac{1}{\Delta E_1 + \Delta E_2}.
\nonumber
\end{eqnarray}
The last line is valid in the limit of narrow shells, where, again, it is understood that the hole wave functions denote the wave functions for the respective shells. For $a_1=
2\, {\rm nm}$, $a_2 = 3\, {\rm nm}$, the characteristic hole level
spacing is of order $10 \, {\rm meV}$. For $\epsilon = 9$, the characteristic energy scale for $|\hbar J_{z,vW}|$ is $ 10 \, {\rm meV}$, significantly larger than the
electrostatic quadrupolar energy derived above.

\subsection{Superconducting charge qubits}
\label{sec:mic-cpbs}

Extending the model of Ref.~\onlinecite{blais:04}, we assume that two small superconducting islands in the
charge regime (so-called Cooper-pair boxes) are located at the center of a resonator and are separated by a
distance $r$, see Fig.~\ref{Fig3} (d). In the charge basis, the total capacitive coupling of the two singly charged Cooper-pair boxes is given by a $z$-$z$
coupling term with energy\cite{marquardt:01,pashkin:03}
\begin{equation}
E_{cc} =\frac{(2e)^2C_c}{(C^{(1)}_\Sigma + C_c)(C^{(2)}_\Sigma + C_c)-C^2_c}.\label{eq:capacitive-energy}
\end{equation}
Here, 
$C_c$ is the capacitance that couples the two islands and $C^{(i)}_\Sigma =
C^{(i)}_R + C^{(i)}_J$ is the total capacitance of the single superconducting island $i=1,2$, where $C^{(i)}_J$
($C^{(i)}_R$) is the capacitance of the island and the corresponding Josephson junction
(the resonator).~\cite{blais:04,wallraff:04}  From the recent experiments using a single Cooper-pair box in a
resonator~\cite{wallraff:04,schuster:05} we infer 
$C^{(i)}_\Sigma \approx e^2 / 40 \mu \mathrm{eV} \approx 4\;\mathrm{fF}$.
We give a rough estimate for $C_c$ in the following and approximate the
Cooper pair boxes as two rectangular plates of length $l$ and width $w$.
The electric field along the perpendicular bisector of a wire of length $l$ with charge density $\sigma$ is
calculated as $E_{\mathrm{wire}}(x)=\sigma l/4 \pi \varepsilon \varepsilon_0 x \sqrt{x^2+l^2/4}$, where $x$ is
the distance from the wire. Integrating over the width $w$ we obtain
\begin{eqnarray}
E_{\mathrm{plate}}(x) & = &  \frac{Q}{2 \pi \varepsilon \varepsilon_0 lw}
\left[\mathrm{arcsinh}\left(\frac{l}{2x}\right) \right.\nonumber\\
 & & \quad \left.- \mathrm{arcsinh}\left(\frac{l}{2(x+w)}\right)\right]
\end{eqnarray}
 for the central electric field at a distance $x$ from the border of the plate [see Fig.~\ref{Fig3} (d)], where $Q=\sigma lw$ is the charge on
 the island.
We obtain for the capacity of the two islands
\begin{eqnarray}
C_c & = & Q \left[2\int_0^{r}{d}x E_{\mathrm{plate}}(x)\right]^{-1}
\nonumber\\
 & = &  \frac{2 \pi \varepsilon \varepsilon_0 w}{K\left(\frac{2r}{l}\right)+K\left(\frac{2w}{l}\right)-
 K\left(\frac{2(w+r)}{l}\right)},
\end{eqnarray}
where we use $K(x)=x\,\mathrm{arcsinh}(1/x)+\mathrm{arcsinh}(x)$.
For $l = 0.1 \;\mu\mathrm{m}$, $w = 1 \;\mu\mathrm{m}$, and $r = 0.5 \;\mu\mathrm{m}$, we
thus obtain $C_c \approx 15.5 \varepsilon \; \mathrm{aF}$, where $\varepsilon$ should be understood 
as the average dielectric constant defined for the space between the Cooper pair boxes. 
For $\varepsilon = 9$, $C_c \approx 0.14 \; \mathrm{fF}$.

Using the qubit eigenstates at the charge degeneracy point, $|\!\uparrow \rangle$, $|\!\downarrow\rangle$, the capacitive interqubit coupling translates into
$\hbar J_x \hat{s}_{1,x}\hat{s}_{2,x}$ with $\hbar J_x = E_{cc}$ [Eq.~(\ref{eq:capacitive-energy})]. 
This interaction mixes the states $|\!\uparrow\uparrow\rangle$ and
$|\!\downarrow\downarrow\rangle$ with the mixing angle $\mathrm{arctan} (J_x/[2(\Omega_1 + \Omega_2)])$. Since the single-qubit splittings $\Omega_i$ are given by the Josephson energy ($\approx 30 \;\mu \mathrm{eV}$ in Ref.~\onlinecite{wallraff:04}),  $ J_x \ll 2(\Omega_1 + \Omega_2)$ and the interqubit coupling may be approximated by $(\hbar J_x /4)( \hat{s}_{1}^+\hat{s}_{2}^- +
\hat{s}_{1}^-\hat{s}_{2}^+)$.
Therefore, capacitively coupled qubits interacting with a cavity may be mapped onto Eq.~(\ref{eq:gen-ham}), wherein $J_\perp = J_x/2$ and $J_z = 0$. 
 For $C_c \approx  0.14 \; \mathrm{fF}$ and characteristic parameters for Cooper pair boxes in a resonator,\cite{wallraff:04} we obtain for the coupling energy
$\hbar J_\perp = E_{cc}/2 \approx 2.6 \; \mu \mathrm{eV}$.

\section{Conclusion}
\label{sec:conclusion}
We have studied two directly coupled qubits which interact perturbatively with a single off-resonant cavity mode. 
Based on the Hamiltonian Eq.~(\ref{eq:gen-ham}) we have provided a diagonalization by means of 
a Schrieffer-Wolff transformation for various cases in order to obtain an effective Hamiltonian describing the perturbative interaction of the cavity with the two-qubit system. 
For identical qubits, the effective Hamiltonian is given by Eq.~(\ref{eq:gen-ham-transf}), 
while for the more general case of non-identical qubits the effective Hamiltonian is given by Eqs.~(\ref{eq:gen-ham-transf2}) -- (\ref{eq:gen-ham-transf2c}). 
Beyond the RWA for the optical interaction, the effective Hamiltonian for non-identical qubits has been obtained as Eqs.~(\ref{eq:gen-ham-transf3}) -- (\ref{eq:gen-ham-transf3c}).  
Based on these results, we have shown that the state of the coupled qubits can be read out from the photon dispersion, the coupled qubit state being projected onto a singlet-triplet basis.
Further, we have investigated the effect of the direct qubit interaction on a cavity-mediated $\sqrt{i\mathrm{SWAP}}$ gate and have shown that a two-qubit gate can also be realized with the total qubit-qubit interaction for a finite residual direct inter-qubit coupling which might be present for, e.g., Cooper-pair boxes in a resonator.
For asymmetric coupling of the qubits to the cavity, we have discussed the control of the direct qubit interaction via the ac Stark effect. 
Finally, we have derived explicit expressions for the energy scales of the Hamiltonian for three different qubit systems, namely quantum dots, quantum shells, and Cooper-pair boxes, and we have calculated the microscopic coupling constants $J_\perp$ and $J_z$ of the direct qubit interaction.

\section*{Acknowledgements}
We thank R. Hanson and W.~H. Lau for discussions and acknowledge support from DARPA SPINS, DARPA QUIST, CNID, ARO, ONR, NCCR Nanoscience, and the Swiss NSF. 

\appendix

\section{Strongly asymmetric qubit-field coupling}
\label{sec:asym}

In order to clarify the physical processes that correspond to
$\hat{H}_{\rm mix}$ in Eq.~(\ref{eq:gen-ham-transf2c}), we consider
the case of two qubits with identical level spacing,
$\Omega_1 = \Omega_2 = \Omega$, but with strongly asymmetric
qubit-cavity mode coupling, $g_1 = g \neq 0$, $g_2=0$.
As will be discussed in more detail below, $\hat{H}_{\rm mix}$
then describes the ${\it competition}$ between the direct
coupling and the cavity-induced Stark and Lamb shift.
For strongly asymmetric coupling, only qubit $1$ will
experience a Stark and Lamb shift and hence will be shifted out of
resonance with qubit $2$. In particular, if the Stark shift of
qubit $1$ is large compared to $J_{\perp}$, the qubits will become
effectively decoupled.

In order to quantify our argument, we evaluate the terms proportional
to $|\psi_{s,a}\rangle\langle\psi_{s,a}|$
in Eqs.~(\ref{eq:gen-ham-transf2a}) and (\ref{eq:gen-ham-transf2c})
in the limit of weak direct inter-qubit coupling,
$|J_\perp| \ll |\omega - (\Omega \pm J_z/2)|$.
Then the denominators in the expressions for $\tilde{\gamma}_i$
[Eq.~(\ref{eq:swtrafo5})] can be replaced by
$E_{s,a} - E_- \rightarrow \omega - (\Omega - J_z/2)$ and  by
$E_+-E_{s,a} \rightarrow \omega - (\Omega + J_z/2)$. Then,
\begin{eqnarray}
&&\hat{H}_{\rm S,L} + \hat{H}_{\rm mix} =
-\hbar  g^2 \frac{\left(\hat{a}^\dagger \hat{a} + 1\right)|T_+\rangle\langle T_+|}{\omega-(\Omega+J_z/2) }\label{eq:asym-exp1} \\
&& \hspace*{0.5cm}  +\hbar  g^2
\frac{\hat{a}^\dagger \hat{a}|T_-\rangle\langle T_-|}{\omega-(\Omega-J_z/2) }
-\hbar  g^2 \frac{\left(\hat{a}^\dagger \hat{a} + 1\right)
|\uparrow\downarrow\rangle\langle \uparrow\downarrow|}{\omega-(\Omega-J_z/2)}
\nonumber  \\
&&  \hspace*{0.5cm} + \hbar  g^2 \frac{\hat{a}^\dagger \hat{a}
|\downarrow \uparrow\rangle\langle \downarrow \uparrow|}{\omega-(\Omega+J_z/2)},
\nonumber
\end{eqnarray}
which represents the Stark and Lamb shift of qubit $1$. In particular, the last two terms in
Eq.~(\ref{eq:asym-exp1}) represent the relative energy shift of $|\uparrow\downarrow\rangle$ and $|\downarrow
\uparrow\rangle$ induced by the cavity. As discussed in Sec.~\ref{sec:disp-control}, this energy shift can be
utilized to tune the qubits into and out of resonance by controlling the photon number in the cavity mode.

\section{Schrieffer-Wolff transformation for general Hamiltonian without RWA}
\label{sec:sw-general2}
For the coupling of the two-qubit system to the cavity
via linearly polarized transitions, the full Hamiltonian reads in general
\begin{eqnarray}
\hat{H}_{\mathrm{full}} &=& \hbar  \sum_{i=1,2} \left[ \Omega_i \hat{s}_{i,z} +
g_i \left(\hat{a}  + \hat{a}^\dagger  \right)
\left(\hat{s}_i^+ + \hat{s}_i^- \right)\right] \label{eq:gen-ham-full} \\
&& + \hbar \omega \hat{a}^\dagger \hat{a} + \frac{\hbar
J_\perp}{2} \left( \hat{s}_1^+  \hat{s}_2^- + \hat{s}_1^-
\hat{s}_2^+ \right) + \hbar J_z \hat{s}_{1,z} \hat{s}_{2,z}.
\nonumber
\end{eqnarray}
The energy non-conserving terms are usually neglected
because of their smallness, leading to the Hamiltonian Eq.~(\ref{eq:gen-ham}). 
In this Appendix we analyze the additional contribution of the
energy non-conserving terms to the coupled-qubit dynamics.

As shown in Section \ref{sec:sw-general}, the eigenstates of the coupled qubits are given by
Eq.~(\ref{eq:gen-es}) for vanishing qubit-cavity coupling. In this basis,
\begin{eqnarray}
\hat{H}_{\mathrm{full}} &= &\sum_{i=\pm} \hbar E_i |T_i\rangle \langle T_i|
+  \sum_{i=s,a} \hbar  E_i |\psi_i\rangle \langle \psi_i|
+ \hbar \omega \hat{a}^\dagger \hat{a} \nonumber \\
&& +  \hbar (\hat{a}+\hat{a}^\dagger) \left[
\gamma_1 |\psi_s\rangle \langle T_-| + \gamma_2  |\psi_a\rangle \langle T_-|\right.\nonumber\\
& & \hspace*{0.5cm} \left. + \gamma_3 |T_+\rangle \langle \psi_s|
  + \gamma_4 |T_+\rangle \langle \psi_a| +H.c. \right]   \label{eq:gen-ham2full},
\end{eqnarray}
with the coupling constants $\gamma_i$ given by
Eqs.~(\ref{eq:gen-coupling-const}) and (\ref{eq:gen-coupling-const2}).
To first order in $\gamma_i$, the Hamiltonian $\hat{H}_{\mathrm{full}}$ is
diagonalized by the Schrieffer-Wolff transformation
\begin{eqnarray}
\hat{H}^\prime_{\mathrm{full}} & = &
e^{\hat{a}(\hat{B} - \hat{B}_{+}^\dagger)
- \hat{a}^\dagger ( \hat{B}^\dagger -  \hat{B}_+ )} \nonumber \\
& & \times \hat{H}_{\mathrm{full}}
e^{-\hat{a}(\hat{B} - \hat{B}_{+}^\dagger)
+ \hat{a}^\dagger ( \hat{B}^\dagger -  \hat{B}_+) }, \label{eq:swtrafo6}
\end{eqnarray}
where $\hat{B}$ is defined by Eq.~(\ref{eq:swtrafo4}) and
\begin{eqnarray}
\hat{B}_+ & = & \tilde{\gamma}_1^+ |\psi_s\rangle \langle T_-| +  \tilde{\gamma}_2^+ |\psi_a\rangle \langle T_-| \label{eq:swtrafo7} \\
&& +  \tilde{\gamma}_3^+ |T_+\rangle \langle \psi_s| +\tilde{\gamma}_4^+ |T_+\rangle \langle \psi_a|, \nonumber
\end{eqnarray}
with the coefficients
\begin{eqnarray}
\tilde{\gamma}_1^+ & = & \frac{\gamma_1}{\omega + E_s - E_- },\nonumber \\
\tilde{\gamma}_2^+ & = & \frac{\gamma_2}{\omega + E_a - E_- }, \label{eq:swtrafo8} \\
\tilde{\gamma}_3^+ & = & \frac{\gamma_3}{\omega + E_+ - E_s }, \nonumber \\
\tilde{\gamma}_4^+ & = & \frac{\gamma_4}{\omega + E_+ - E_a }. \nonumber
\end{eqnarray}
Clearly, $|\tilde{\gamma}_i^+ |< |\tilde{\gamma}_i|$ for all $i$ and $\omega >0$. Yet,
$\tilde{\gamma}_i^+ \sim \tilde{\gamma}_i$ for $\omega \ll E_n - E_m$, and $|\tilde{\gamma}_i^+| \sim |\tilde{\gamma}_i|$ for $\omega \gg E_n - E_m$, where $n \neq m$ and $n,m \in \{s,a,+,-\}$. 

Instead of
Eq.~(\ref{eq:gen-ham-transf2}), we obtain to second order in $\gamma_i$ the effective Hamiltonian,
\begin{equation}
\hat{H}^\prime_{\mathrm{full}} = \hat{H}_0 + \hat{H}_{\rm S,L}^{\mathrm{full}}
+ \hat{H}_{\rm 2-ph}^{\mathrm{full}} +  \hat{H}_{\rm mix}^{\mathrm{full}}.
\label{eq:gen-ham-transf3}
\end{equation}

Here, $\hat{H}_0$ is given by Eq.~(\ref{eq:h0}). The term for the Stark- and Lamb-shift
is now calculated to be

\begin{widetext}
\begin{eqnarray}
\hat{H}_{\rm S,L}^{\mathrm{full}} & = &
\frac{\hbar}{2}(\hat{a}^2  + \hat{a}^{\dagger 2} + 2 \hat{a}^\dagger\hat{a} + 2) \label{eq:starkfull}\\
 & & \times \left\{
\left[\gamma_1(\tilde{\gamma}_1 + \tilde{\gamma}_1^+ )
- \gamma_3 (\tilde{\gamma}_3 + \tilde{\gamma}_3^+ )\right]
|\psi_s \rangle \langle \psi_s| + \left[\gamma_2(\tilde{\gamma}_2 + \tilde{\gamma}_2^+ )
- \gamma_4 (\tilde{\gamma}_4 + \tilde{\gamma}_4^+ )\right] |\psi_a \rangle \langle \psi_a|\right.  \nonumber \\
 & & \left.+ \left[\gamma_3(\tilde{\gamma}_3 + \tilde{\gamma}_3^+ )+\gamma_4 (\tilde{\gamma}_4 + \tilde{\gamma}_4^+) \right] |T_+ \rangle \langle T_+|-\left[\gamma_1 (\tilde{\gamma}_1+ \tilde{\gamma}_1^+ )
+ \gamma_2  (\tilde{\gamma}_2 + \tilde{\gamma}_2^+) \right] |T_- \rangle \langle T_-| \right\}\nonumber\\
& & + \hbar\left[\gamma_3\tilde{\gamma}_3 |\psi_s \rangle \langle \psi_s|
+\gamma_4\tilde{\gamma}_4 |\psi_a \rangle \langle \psi_a| + \left(\gamma_1\tilde{\gamma}_1 + \gamma_2\tilde{\gamma}_2\right)
|T_- \rangle \langle T_-| \right]. \nonumber 
\end{eqnarray}
\end{widetext}
In contrast to the result using the RWA, the additional terms containing $\gamma_i^+$ modify the Stark shift, see Fig.~\ref{FigStark}, and also the Lamb shift. In Fig.~\ref{FigStark} we have assumed a photon Fock state in the cavity. 
Further, the two-photon transition term is given by
\begin{widetext}
\begin{eqnarray}
\hat{H}_{\rm 2-ph}^{\mathrm{full}} & = & \frac{\hbar}{2} \left[ \left(\tilde{\gamma}_3 \gamma_1 -
\tilde{\gamma}_1 \gamma_3 + \tilde{\gamma}_4\gamma_2
- \tilde{\gamma}_2 \gamma_4 \right)\left(\hat{a}^2 |T_+\rangle \langle T_-| + \hat{a}^{\dagger 2} |T_-\rangle \langle T_+| \right)\right.   \\
 & & + \left(\tilde{\gamma}_3^+ \gamma_1 - \tilde{\gamma}_1^+ \gamma_3 +
\tilde{\gamma}_4^+\gamma_2
- \tilde{\gamma}_2^+ \gamma_4 \right)  \left(\hat{a}^{\dagger 2} |T_+\rangle \langle T_-| + \hat{a}^{2} |T_-\rangle \langle T_+| \right)\nonumber \\
 & & + \left(\tilde{\gamma}_2 \gamma_1 - \tilde{\gamma}_3 \gamma_4 +
\tilde{\gamma}_1^+\gamma_2 - \tilde{\gamma}_4^+ \gamma_3 \right)
 \left(\hat{a}^{\dagger 2} |\psi_s\rangle \langle \psi_a|
+ \hat{a}^{2} |\psi_a\rangle \langle \psi_s| \right)\nonumber \\
&&  \left. + \left(\tilde{\gamma}_1 \gamma_2 - \tilde{\gamma}_4 \gamma_3 + \tilde{\gamma}_2^+\gamma_1 -
\tilde{\gamma}_3^+ \gamma_4 \right)
 \left(\hat{a}^{ 2} |\psi_s\rangle \langle \psi_a|
+ \hat{a}^{\dagger 2} |\psi_a\rangle \langle \psi_s| \right)  \right], \nonumber
\end{eqnarray}
\end{widetext}
which consists of $\hat{H}_{\mathrm{2-ph}}$ and additional energy non-conserving
transitions between $|T_+\rangle$ and $|T_-\rangle$ as well as between $|\psi_s\rangle$ and $|\psi_a\rangle$.
Similarly,
\begin{widetext}
\begin{eqnarray}
\hat{H}_{\rm mix}^{\mathrm{full}} & = & \frac{\hbar}{2}
\{
\left[\gamma_1(\tilde{\gamma}_2 + \tilde{\gamma}_2^+ )
+ \gamma_2(\tilde{\gamma}_1 + \tilde{\gamma}_1^+ )  - \gamma_3 (\tilde{\gamma}_4 + \tilde{\gamma}_4^+ )
- \gamma_4 (\tilde{\gamma}_3 + \tilde{\gamma}_3^+ )\right]  (\hat{a}^\dagger\hat{a} + 1)
(|\psi_s\rangle \langle \psi_a| + |\psi_a\rangle \langle \psi_s|)  \label{eq:gen-ham-transf3c}\\
&& +\left(\gamma_1\tilde{\gamma}_2^+ - \gamma_2\tilde{\gamma}_1^+
+ \gamma_3\tilde{\gamma}_4 + \gamma_4\tilde{\gamma}_3\right)
|\psi_s\rangle \langle \psi_a| +\left(\gamma_2\tilde{\gamma}_1^+ -
\gamma_1\tilde{\gamma}_2^+
+ \gamma_3\tilde{\gamma}_4 + \gamma_4\tilde{\gamma}_3\right)
|\psi_a\rangle \langle \psi_s| \nonumber \\
&& +\left[\gamma_1(\tilde{\gamma}_3 + \tilde{\gamma}_3^+ )
+ \gamma_2(\tilde{\gamma}_4 + \tilde{\gamma}_4^+ )  - \gamma_3 (\tilde{\gamma}_1 + \tilde{\gamma}_1^+ )
- \gamma_4 (\tilde{\gamma}_2 + \tilde{\gamma}_2^+ )\right] (\hat{a}^\dagger\hat{a} + 1)
(|T_+\rangle \langle T_-| + |T_-\rangle \langle T_+|) \nonumber\\
&& +\left(\gamma_3\tilde{\gamma}_1 + \gamma_4\tilde{\gamma}_2
- \gamma_1\tilde{\gamma}_3^+ - \gamma_2\tilde{\gamma}_4^+\right)
|T_+\rangle \langle T_-| +\left(\gamma_1\tilde{\gamma}_3^+ + \gamma_3\tilde{\gamma}_1
+ \gamma_2\tilde{\gamma}_4^+ + \gamma_4\tilde{\gamma}_2\right)
|T_-\rangle \langle T_+|\nonumber \}
\end{eqnarray}
\end{widetext}
contains $\hat{H}_{\mathrm{mix}}$ and new terms that couple
$|T_+\rangle$ and $|T_-\rangle$ in addition to $|\psi_s\rangle$ and
$|\psi_a\rangle$. In contrast to Eq.~(\ref{eq:gen-ham-transf2c}), the above term induces mixing of $|T_+\rangle$ and $|T_-\rangle$ even for identical qubits.

%\bibliography{qdmolecule}

\begin{thebibliography}{39}
\expandafter\ifx\csname natexlab\endcsname\relax\def\natexlab#1{#1}\fi
\expandafter\ifx\csname bibnamefont\endcsname\relax
  \def\bibnamefont#1{#1}\fi
\expandafter\ifx\csname bibfnamefont\endcsname\relax
  \def\bibfnamefont#1{#1}\fi
\expandafter\ifx\csname citenamefont\endcsname\relax
  \def\citenamefont#1{#1}\fi
\expandafter\ifx\csname url\endcsname\relax
  \def\url#1{\texttt{#1}}\fi
\expandafter\ifx\csname urlprefix\endcsname\relax\def\urlprefix{URL }\fi
\providecommand{\bibinfo}[2]{#2}
\providecommand{\eprint}[2][]{\url{#2}}

\bibitem[{\citenamefont{Raimond et~al.}(2001)\citenamefont{Raimond, Brune, and
  Haroche}}]{raimond:01}
\bibinfo{author}{\bibfnamefont{J.~M.} \bibnamefont{Raimond}},
  \bibinfo{author}{\bibfnamefont{M.}~\bibnamefont{Brune}}, \bibnamefont{and}
  \bibinfo{author}{\bibfnamefont{S.}~\bibnamefont{Haroche}},
  \bibinfo{journal}{Rev. Mod. Phys.} \textbf{\bibinfo{volume}{73}},
  \bibinfo{pages}{565} (\bibinfo{year}{2001}).

\bibitem[{\citenamefont{Turchette et~al.}(1995)\citenamefont{Turchette, Hood,
  Lange, Mabuchi, and Kimble}}]{turchette:95}
\bibinfo{author}{\bibfnamefont{Q.~A.} \bibnamefont{Turchette}},
  \bibinfo{author}{\bibfnamefont{C.~J.} \bibnamefont{Hood}},
  \bibinfo{author}{\bibfnamefont{W.}~\bibnamefont{Lange}},
  \bibinfo{author}{\bibfnamefont{H.}~\bibnamefont{Mabuchi}}, \bibnamefont{and}
  \bibinfo{author}{\bibfnamefont{H.~J.} \bibnamefont{Kimble}},
  \bibinfo{journal}{Phys. Rev. Lett.} \textbf{\bibinfo{volume}{75}},
  \bibinfo{pages}{4710} (\bibinfo{year}{1995}).

\bibitem[{\citenamefont{\mbox{B. T. H.} Varcoe
  et~al.}(2000)\citenamefont{\mbox{B. T. H.} Varcoe, Brattke, Weidinger, and
  Walther}}]{varcoe:00}
\bibinfo{author}{\bibnamefont{\mbox{B. T. H.} Varcoe}},
  \bibinfo{author}{\bibfnamefont{S.}~\bibnamefont{Brattke}},
  \bibinfo{author}{\bibfnamefont{M.}~\bibnamefont{Weidinger}},
  \bibnamefont{and} \bibinfo{author}{\bibfnamefont{H.}~\bibnamefont{Walther}},
  \bibinfo{journal}{Nature} \textbf{\bibinfo{volume}{403}},
  \bibinfo{pages}{743} (\bibinfo{year}{2000}).

\bibitem[{\citenamefont{van Enk et~al.}(1997)\citenamefont{van Enk, Cirac, and
  Zoller}}]{enk:97}
\bibinfo{author}{\bibfnamefont{S.~J.} \bibnamefont{van Enk}},
  \bibinfo{author}{\bibfnamefont{J.~I.} \bibnamefont{Cirac}}, \bibnamefont{and}
  \bibinfo{author}{\bibfnamefont{P.}~\bibnamefont{Zoller}},
  \bibinfo{journal}{Phys. Rev. Lett.} \textbf{\bibinfo{volume}{78}},
  \bibinfo{pages}{4293} (\bibinfo{year}{1997}).

\bibitem[{\citenamefont{Blais et~al.}(2004)\citenamefont{Blais, Huang,
  Wallraff, Girvin, and Schoelkopf}}]{blais:04}
\bibinfo{author}{\bibfnamefont{A.}~\bibnamefont{Blais}},
  \bibinfo{author}{\bibfnamefont{R.-S.} \bibnamefont{Huang}},
  \bibinfo{author}{\bibfnamefont{A.}~\bibnamefont{Wallraff}},
  \bibinfo{author}{\bibfnamefont{S.~M.} \bibnamefont{Girvin}},
  \bibnamefont{and} \bibinfo{author}{\bibfnamefont{R.~J.}
  \bibnamefont{Schoelkopf}}, \bibinfo{journal}{Phys. Rev. A}
  \textbf{\bibinfo{volume}{69}}, \bibinfo{pages}{062320}
  (\bibinfo{year}{2004}).

\bibitem[{\citenamefont{Wallraff et~al.}(2004)\citenamefont{Wallraff, Schuster,
  Blais, Frunzio, Huang, Majer, Kumar, Girvin, and Schoelkopf}}]{wallraff:04}
\bibinfo{author}{\bibfnamefont{A.}~\bibnamefont{Wallraff}},
  \bibinfo{author}{\bibfnamefont{D.~I.} \bibnamefont{Schuster}},
  \bibinfo{author}{\bibfnamefont{A.}~\bibnamefont{Blais}},
  \bibinfo{author}{\bibfnamefont{L.}~\bibnamefont{Frunzio}},
  \bibinfo{author}{\bibfnamefont{R.-S.} \bibnamefont{Huang}},
  \bibinfo{author}{\bibfnamefont{J.}~\bibnamefont{Majer}},
  \bibinfo{author}{\bibfnamefont{S.}~\bibnamefont{Kumar}},
  \bibinfo{author}{\bibfnamefont{S.~M.} \bibnamefont{Girvin}},
  \bibnamefont{and} \bibinfo{author}{\bibfnamefont{R.~J.}
  \bibnamefont{Schoelkopf}}, \bibinfo{journal}{Nature}
  \textbf{\bibinfo{volume}{431}}, \bibinfo{pages}{162} (\bibinfo{year}{2004}).

\bibitem[{\citenamefont{Imamo\={g}lu}(2000)}]{imamoglu:00}
\bibinfo{author}{\bibfnamefont{A.}~\bibnamefont{Imamo\={g}lu}},
  \bibinfo{journal}{Fortschr. Phys.} \textbf{\bibinfo{volume}{48}},
  \bibinfo{pages}{987} (\bibinfo{year}{2000}).

\bibitem[{\citenamefont{Imamo\={g}lu et~al.}(1999)\citenamefont{Imamo\={g}lu,
  Awschalom, Burkard, DiVincenzo, Loss, Sherwin, and Small}}]{imamoglu:99}
\bibinfo{author}{\bibfnamefont{A.}~\bibnamefont{Imamo\={g}lu}},
  \bibinfo{author}{\bibfnamefont{D.~D.} \bibnamefont{Awschalom}},
  \bibinfo{author}{\bibfnamefont{G.}~\bibnamefont{Burkard}},
  \bibinfo{author}{\bibfnamefont{D.~P.} \bibnamefont{DiVincenzo}},
  \bibinfo{author}{\bibfnamefont{D.}~\bibnamefont{Loss}},
  \bibinfo{author}{\bibfnamefont{M.}~\bibnamefont{Sherwin}}, \bibnamefont{and}
  \bibinfo{author}{\bibfnamefont{A.}~\bibnamefont{Small}},
  \bibinfo{journal}{Phys. Rev. Lett.} \textbf{\bibinfo{volume}{83}},
  \bibinfo{pages}{4204} (\bibinfo{year}{1999}).

\bibitem[{\citenamefont{Kiraz et~al.}(2003)\citenamefont{Kiraz, Reese, Gayral,
  Zhang, Schoenfeld, Gerardot, Petroff, Hu, and Imamo\={g}lu}}]{kiraz:03}
\bibinfo{author}{\bibfnamefont{A.}~\bibnamefont{Kiraz}},
  \bibinfo{author}{\bibfnamefont{C.}~\bibnamefont{Reese}},
  \bibinfo{author}{\bibfnamefont{B.}~\bibnamefont{Gayral}},
  \bibinfo{author}{\bibfnamefont{L.}~\bibnamefont{Zhang}},
  \bibinfo{author}{\bibfnamefont{W.~V.} \bibnamefont{Schoenfeld}},
  \bibinfo{author}{\bibfnamefont{B.~D.} \bibnamefont{Gerardot}},
  \bibinfo{author}{\bibfnamefont{P.~M.} \bibnamefont{Petroff}},
  \bibinfo{author}{\bibfnamefont{E.~L.} \bibnamefont{Hu}}, \bibnamefont{and}
  \bibinfo{author}{\bibfnamefont{A.}~\bibnamefont{Imamo\={g}lu}},
  \bibinfo{journal}{J. Opt. B} \textbf{\bibinfo{volume}{5}},
  \bibinfo{pages}{129} (\bibinfo{year}{2003}).

\bibitem[{\citenamefont{Vu\v{c}kovi\'{c} and Yamamoto}(2003)}]{vuckovic:03}
\bibinfo{author}{\bibfnamefont{J.}~\bibnamefont{Vu\v{c}kovi\'{c}}}
  \bibnamefont{and} \bibinfo{author}{\bibfnamefont{Y.}~\bibnamefont{Yamamoto}},
  \bibinfo{journal}{Appl. Phys. Lett.} \textbf{\bibinfo{volume}{82}},
  \bibinfo{pages}{2374} (\bibinfo{year}{2003}).

\bibitem[{\citenamefont{Loss and DiVincenzo}(1998)}]{loss:98}
\bibinfo{author}{\bibfnamefont{D.}~\bibnamefont{Loss}} \bibnamefont{and}
  \bibinfo{author}{\bibfnamefont{D.~P.} \bibnamefont{DiVincenzo}},
  \bibinfo{journal}{Phys. Rev. A} \textbf{\bibinfo{volume}{57}},
  \bibinfo{pages}{120} (\bibinfo{year}{1998}).

\bibitem[{\citenamefont{Burkard et~al.}(1999)\citenamefont{Burkard, Loss, and
  DiVincenzo}}]{burkard:99}
\bibinfo{author}{\bibfnamefont{G.}~\bibnamefont{Burkard}},
  \bibinfo{author}{\bibfnamefont{D.}~\bibnamefont{Loss}}, \bibnamefont{and}
  \bibinfo{author}{\bibfnamefont{D.~P.} \bibnamefont{DiVincenzo}},
  \bibinfo{journal}{Phys. Rev. B} \textbf{\bibinfo{volume}{59}},
  \bibinfo{pages}{2070} (\bibinfo{year}{1999}).

\bibitem[{\citenamefont{Burkard et~al.}(2000)\citenamefont{Burkard, Seelig, and
  Loss}}]{burkard:00}
\bibinfo{author}{\bibfnamefont{G.}~\bibnamefont{Burkard}},
  \bibinfo{author}{\bibfnamefont{G.}~\bibnamefont{Seelig}}, \bibnamefont{and}
  \bibinfo{author}{\bibfnamefont{D.}~\bibnamefont{Loss}},
  \bibinfo{journal}{Phys. Rev. B} \textbf{\bibinfo{volume}{62}},
  \bibinfo{pages}{2581} (\bibinfo{year}{2000}).

\bibitem[{\citenamefont{Dicke}(1954)}]{dicke:54}
\bibinfo{author}{\bibfnamefont{R.~H.} \bibnamefont{Dicke}},
  \bibinfo{journal}{Phys. Rev.} \textbf{\bibinfo{volume}{93}},
  \bibinfo{pages}{99} (\bibinfo{year}{1954}).

\bibitem[{\citenamefont{Schuster et~al.}(2005)\citenamefont{Schuster, Wallraff,
  Blais, Frunzio, Huang, Majer, Girvin, and Schoelkopf}}]{schuster:05}
\bibinfo{author}{\bibfnamefont{D.~I.} \bibnamefont{Schuster}},
  \bibinfo{author}{\bibfnamefont{A.}~\bibnamefont{Wallraff}},
  \bibinfo{author}{\bibfnamefont{A.}~\bibnamefont{Blais}},
  \bibinfo{author}{\bibfnamefont{L.}~\bibnamefont{Frunzio}},
  \bibinfo{author}{\bibfnamefont{R.-S.} \bibnamefont{Huang}},
  \bibinfo{author}{\bibfnamefont{J.}~\bibnamefont{Majer}},
  \bibinfo{author}{\bibfnamefont{S.~M.} \bibnamefont{Girvin}},
  \bibnamefont{and} \bibinfo{author}{\bibfnamefont{R.~J.}
  \bibnamefont{Schoelkopf}}, \bibinfo{journal}{Phys. Rev. Lett.}
  \textbf{\bibinfo{volume}{94}}, \bibinfo{pages}{123602}
  (\bibinfo{year}{2005}).

\bibitem[{\citenamefont{Hastings et~al.}(2005)\citenamefont{Hastings, de~Dood,
  Kim, Marshall, Eisenberg, and Bouwmeester}}]{hastings:05}
\bibinfo{author}{\bibfnamefont{S.}~\bibnamefont{Hastings}},
  \bibinfo{author}{\bibfnamefont{M.~J.~A.} \bibnamefont{de~Dood}},
  \bibinfo{author}{\bibfnamefont{H.}~\bibnamefont{Kim}},
  \bibinfo{author}{\bibfnamefont{W.}~\bibnamefont{Marshall}},
  \bibinfo{author}{\bibfnamefont{H.~S.} \bibnamefont{Eisenberg}},
  \bibnamefont{and}
  \bibinfo{author}{\bibfnamefont{D.}~\bibnamefont{Bouwmeester}},
  \bibinfo{journal}{Appl. Phys. Lett.} \textbf{\bibinfo{volume}{86}},
  \bibinfo{pages}{031109} (\bibinfo{year}{2005}).

\bibitem[{\citenamefont{Unold et~al.}(2004)\citenamefont{Unold, Mueller,
  Lienau, Elsaesser, and Wieck}}]{unold:04}
\bibinfo{author}{\bibfnamefont{T.}~\bibnamefont{Unold}},
  \bibinfo{author}{\bibfnamefont{K.}~\bibnamefont{Mueller}},
  \bibinfo{author}{\bibfnamefont{C.}~\bibnamefont{Lienau}},
  \bibinfo{author}{\bibfnamefont{T.}~\bibnamefont{Elsaesser}},
  \bibnamefont{and} \bibinfo{author}{\bibfnamefont{A.~D.} \bibnamefont{Wieck}},
  \bibinfo{journal}{Phys. Rev. Lett.} \textbf{\bibinfo{volume}{92}},
  \bibinfo{pages}{157401} (\bibinfo{year}{2004}).

\bibitem[{\citenamefont{Badolato et~al.}(2005)\citenamefont{Badolato, Hennessy,
  Atat\mbox{\"u}re, Dreiser, Hu, Petroff, and Imamo\v{g}lu}}]{badolato:05}
\bibinfo{author}{\bibfnamefont{A.}~\bibnamefont{Badolato}},
  \bibinfo{author}{\bibfnamefont{K.}~\bibnamefont{Hennessy}},
  \bibinfo{author}{\bibfnamefont{M.}~\bibnamefont{Atat\mbox{\"u}re}},
  \bibinfo{author}{\bibfnamefont{J.}~\bibnamefont{Dreiser}},
  \bibinfo{author}{\bibfnamefont{E.}~\bibnamefont{Hu}},
  \bibinfo{author}{\bibfnamefont{P.~M.} \bibnamefont{Petroff}},
  \bibnamefont{and}
  \bibinfo{author}{\bibfnamefont{A.}~\bibnamefont{Imamo\v{g}lu}},
  \bibinfo{journal}{Science} \textbf{\bibinfo{volume}{308}},
  \bibinfo{pages}{1158} (\bibinfo{year}{2005}).

\bibitem[{\citenamefont{Nazir et~al.}(2004)\citenamefont{Nazir, Lovett, and
  \mbox{G. A. D.} Briggs}}]{nazir:04}
\bibinfo{author}{\bibfnamefont{A.}~\bibnamefont{Nazir}},
  \bibinfo{author}{\bibfnamefont{B.~W.} \bibnamefont{Lovett}},
  \bibnamefont{and} \bibinfo{author}{\bibnamefont{\mbox{G. A. D.} Briggs}},
  \bibinfo{journal}{Phys. Rev. A} \textbf{\bibinfo{volume}{70}},
  \bibinfo{pages}{052301} (\bibinfo{year}{2004}).

\bibitem[{\citenamefont{Cerletti
  et~al.}(2005{\natexlab{a}})\citenamefont{Cerletti, Coish, Gywat, and
  Loss}}]{cerletti:05a}
\bibinfo{author}{\bibfnamefont{V.}~\bibnamefont{Cerletti}},
  \bibinfo{author}{\bibfnamefont{W.~A.} \bibnamefont{Coish}},
  \bibinfo{author}{\bibfnamefont{O.}~\bibnamefont{Gywat}}, \bibnamefont{and}
  \bibinfo{author}{\bibfnamefont{D.}~\bibnamefont{Loss}},
  \bibinfo{journal}{Nanotechnology} \textbf{\bibinfo{volume}{16}},
  \bibinfo{pages}{R27} (\bibinfo{year}{2005}{\natexlab{a}}).

\bibitem[{\citenamefont{Biolatti et~al.}(2000)\citenamefont{Biolatti, Iotti,
  Zanardi, and Rossi}}]{biolatti:00}
\bibinfo{author}{\bibfnamefont{E.}~\bibnamefont{Biolatti}},
  \bibinfo{author}{\bibfnamefont{R.~C.} \bibnamefont{Iotti}},
  \bibinfo{author}{\bibfnamefont{P.}~\bibnamefont{Zanardi}}, \bibnamefont{and}
  \bibinfo{author}{\bibfnamefont{F.}~\bibnamefont{Rossi}},
  \bibinfo{journal}{Phys. Rev. Lett.} \textbf{\bibinfo{volume}{85}},
  \bibinfo{pages}{5647} (\bibinfo{year}{2000}).

\bibitem[{\citenamefont{Tsitsishvili et~al.}(2003)\citenamefont{Tsitsishvili,
  v.~Baltz, and Kalt}}]{tsitsishvili:03}
\bibinfo{author}{\bibfnamefont{E.}~\bibnamefont{Tsitsishvili}},
  \bibinfo{author}{\bibfnamefont{R.}~\bibnamefont{v.~Baltz}}, \bibnamefont{and}
  \bibinfo{author}{\bibfnamefont{H.}~\bibnamefont{Kalt}},
  \bibinfo{journal}{Phys. Rev. B} \textbf{\bibinfo{volume}{67}},
  \bibinfo{pages}{205330} (\bibinfo{year}{2003}).

\bibitem[{\citenamefont{Cerletti
  et~al.}(2005{\natexlab{b}})\citenamefont{Cerletti, Gywat, and
  Loss}}]{cerletti:05b}
\bibinfo{author}{\bibfnamefont{V.}~\bibnamefont{Cerletti}},
  \bibinfo{author}{\bibfnamefont{O.}~\bibnamefont{Gywat}}, \bibnamefont{and}
  \bibinfo{author}{\bibfnamefont{D.}~\bibnamefont{Loss}},
  \bibinfo{journal}{Phys. Rev. B} \textbf{\bibinfo{volume}{72}},
  \bibinfo{pages}{115316} (\bibinfo{year}{2005}{\natexlab{b}}).

\bibitem[{\citenamefont{Alivisatos}(1996)}]{alivisatos:96}
\bibinfo{author}{\bibfnamefont{A.~P.} \bibnamefont{Alivisatos}},
  \bibinfo{journal}{J. Phys. Chem.} \textbf{\bibinfo{volume}{100}},
  \bibinfo{pages}{13226} (\bibinfo{year}{1996}).

\bibitem[{\citenamefont{Ouyang and Awschalom}(2003)}]{ouyang:03}
\bibinfo{author}{\bibfnamefont{M.}~\bibnamefont{Ouyang}} \bibnamefont{and}
  \bibinfo{author}{\bibfnamefont{D.~D.} \bibnamefont{Awschalom}},
  \bibinfo{journal}{Science} \textbf{\bibinfo{volume}{301}},
  \bibinfo{pages}{1074} (\bibinfo{year}{2003}).

\bibitem[{\citenamefont{Meier et~al.}(2004)\citenamefont{Meier, Cerletti,
  Gywat, Loss, and Awschalom}}]{meier:04}
\bibinfo{author}{\bibfnamefont{F.}~\bibnamefont{Meier}},
  \bibinfo{author}{\bibfnamefont{V.}~\bibnamefont{Cerletti}},
  \bibinfo{author}{\bibfnamefont{O.}~\bibnamefont{Gywat}},
  \bibinfo{author}{\bibfnamefont{D.}~\bibnamefont{Loss}}, \bibnamefont{and}
  \bibinfo{author}{\bibfnamefont{D.~D.} \bibnamefont{Awschalom}},
  \bibinfo{journal}{Phys. Rev. B} \textbf{\bibinfo{volume}{69}},
  \bibinfo{pages}{195315} (\bibinfo{year}{2004}).

\bibitem[{\citenamefont{Schrier and Whaley}(2005)}]{schrier:05}
\bibinfo{author}{\bibfnamefont{J.}~\bibnamefont{Schrier}} \bibnamefont{and}
  \bibinfo{author}{\bibfnamefont{K.~B.} \bibnamefont{Whaley}},
  \bibinfo{journal}{Phys. Rev. B} \textbf{\bibinfo{volume}{72}},
  \bibinfo{pages}{085320} (\bibinfo{year}{2005}).

\bibitem[{\citenamefont{Govorov}(2003)}]{govorov:03}
\bibinfo{author}{\bibfnamefont{A.~O.} \bibnamefont{Govorov}},
  \bibinfo{journal}{Phys. Rev. B} \textbf{\bibinfo{volume}{68}},
  \bibinfo{pages}{075315} (\bibinfo{year}{2003}).

\bibitem[{\citenamefont{Lovett et~al.}(2003)\citenamefont{Lovett, Reina, Nazir,
  and \mbox{G. A. D.} Briggs}}]{lovett:03}
\bibinfo{author}{\bibfnamefont{B.~W.} \bibnamefont{Lovett}},
  \bibinfo{author}{\bibfnamefont{J.~H.} \bibnamefont{Reina}},
  \bibinfo{author}{\bibfnamefont{A.}~\bibnamefont{Nazir}}, \bibnamefont{and}
  \bibinfo{author}{\bibnamefont{\mbox{G. A. D.} Briggs}},
  \bibinfo{journal}{Phys. Rev. B} \textbf{\bibinfo{volume}{68}},
  \bibinfo{pages}{205319} (\bibinfo{year}{2003}).

\bibitem[{\citenamefont{Nazir et~al.}(2005)\citenamefont{Nazir, Lovett,
  Barrett, Reina, and \mbox{G. A. D.} Briggs}}]{nazir:03}
\bibinfo{author}{\bibfnamefont{A.}~\bibnamefont{Nazir}},
  \bibinfo{author}{\bibfnamefont{B.~W.} \bibnamefont{Lovett}},
  \bibinfo{author}{\bibfnamefont{S.~D.} \bibnamefont{Barrett}},
  \bibinfo{author}{\bibfnamefont{J.~H.} \bibnamefont{Reina}}, \bibnamefont{and}
  \bibinfo{author}{\bibnamefont{\mbox{G. A. D.} Briggs}},
  \bibinfo{journal}{Phys. Rev. B} \textbf{\bibinfo{volume}{71}},
  \bibinfo{pages}{045334} (\bibinfo{year}{2005}).

\bibitem[{\citenamefont{\mbox{Al.} L~Efros}(1992)}]{efros:92}
\bibinfo{author}{\bibnamefont{\mbox{Al.} L~Efros}}, \bibinfo{journal}{Phys.
  Rev. B} \textbf{\bibinfo{volume}{46}}, \bibinfo{pages}{7448}
  (\bibinfo{year}{1992}).

\bibitem[{\citenamefont{Xia}(1989)}]{xia:89}
\bibinfo{author}{\bibfnamefont{J.-B.} \bibnamefont{Xia}},
  \bibinfo{journal}{Phys. Rev. B} \textbf{\bibinfo{volume}{40}},
  \bibinfo{pages}{8500} (\bibinfo{year}{1989}).

\bibitem[{\citenamefont{Meier and Awschalom}(2005)}]{meier:04b}
\bibinfo{author}{\bibfnamefont{F.}~\bibnamefont{Meier}} \bibnamefont{and}
  \bibinfo{author}{\bibfnamefont{D.~D.} \bibnamefont{Awschalom}},
  \bibinfo{journal}{Phys. Rev. B} \textbf{\bibinfo{volume}{71}},
  \bibinfo{pages}{205315} (\bibinfo{year}{2005}).

\bibitem[{\citenamefont{Jackson}(1962)}]{jackson}
\bibinfo{author}{\bibfnamefont{J.~D.} \bibnamefont{Jackson}},
  \emph{\bibinfo{title}{Classical Electrodynamics}}
  (\bibinfo{publisher}{Wiley}, \bibinfo{address}{New York},
  \bibinfo{year}{1962}).

\bibitem[{\citenamefont{Battaglia et~al.}(2003)\citenamefont{Battaglia, Li,
  Wang, and Peng}}]{battaglia:03}
\bibinfo{author}{\bibfnamefont{D.}~\bibnamefont{Battaglia}},
  \bibinfo{author}{\bibfnamefont{J.~J.} \bibnamefont{Li}},
  \bibinfo{author}{\bibfnamefont{Y.}~\bibnamefont{Wang}}, \bibnamefont{and}
  \bibinfo{author}{\bibfnamefont{X.}~\bibnamefont{Peng}},
  \bibinfo{journal}{Angew. Chem. Int. Ed.} \textbf{\bibinfo{volume}{42}},
  \bibinfo{pages}{5035} (\bibinfo{year}{2003}).

\bibitem[{\citenamefont{Berezovsky et~al.}(2005)\citenamefont{Berezovsky,
  Ouyang, Meier, Awschalom, Battaglia, and Peng}}]{berezovsky:04}
\bibinfo{author}{\bibfnamefont{J.}~\bibnamefont{Berezovsky}},
  \bibinfo{author}{\bibfnamefont{M.}~\bibnamefont{Ouyang}},
  \bibinfo{author}{\bibfnamefont{F.}~\bibnamefont{Meier}},
  \bibinfo{author}{\bibfnamefont{D.~D.} \bibnamefont{Awschalom}},
  \bibinfo{author}{\bibfnamefont{D.}~\bibnamefont{Battaglia}},
  \bibnamefont{and} \bibinfo{author}{\bibfnamefont{X.}~\bibnamefont{Peng}},
  \bibinfo{journal}{Phys. Rev. B} \textbf{\bibinfo{volume}{71}},
  \bibinfo{pages}{R081309} (\bibinfo{year}{2005}).

\bibitem[{\citenamefont{Meier}()}]{meier:un}
\bibinfo{author}{\bibfnamefont{F.}~\bibnamefont{Meier}},
  \bibinfo{note}{unpublished}.

\bibitem[{\citenamefont{Marquardt and Bruder}(2001)}]{marquardt:01}
\bibinfo{author}{\bibfnamefont{F.}~\bibnamefont{Marquardt}} \bibnamefont{and}
  \bibinfo{author}{\bibfnamefont{C.}~\bibnamefont{Bruder}},
  \bibinfo{journal}{Phys. Rev. B} \textbf{\bibinfo{volume}{63}},
  \bibinfo{pages}{054514} (\bibinfo{year}{2001}).

\bibitem[{\citenamefont{Pashkin et~al.}(2003)\citenamefont{Pashkin, Yamamoto,
  Astafiev, Nakamura, Averin, and Tsai}}]{pashkin:03}
\bibinfo{author}{\bibfnamefont{Y.~A.} \bibnamefont{Pashkin}},
  \bibinfo{author}{\bibfnamefont{T.}~\bibnamefont{Yamamoto}},
  \bibinfo{author}{\bibfnamefont{O.}~\bibnamefont{Astafiev}},
  \bibinfo{author}{\bibfnamefont{Y.}~\bibnamefont{Nakamura}},
  \bibinfo{author}{\bibfnamefont{D.~V.} \bibnamefont{Averin}},
  \bibnamefont{and} \bibinfo{author}{\bibfnamefont{J.~S.} \bibnamefont{Tsai}},
  \bibinfo{journal}{Nature} \textbf{\bibinfo{volume}{421}},
  \bibinfo{pages}{823} (\bibinfo{year}{2003}).

\end{thebibliography}
%\bibliographystyle{prsty}

\end{document}